\documentclass[a4paper,11pt]{article}
\usepackage{jcappub} % for details on the use of the package, please see the JINST-author-manual
\arxivnumber{2604.07824} % Only if you have one
\title{\boldmath Deciphering the IceCube Diffuse Neutrino Observations via AGN Variability}

\author[a]{Caijin Xie}
\author[a]{Zijian Qiu}
\author[a]{Yudong Cui}
\author[b,c]{Sujie Lin}
\author[a,d]{Lili Yang}

\affiliation[a]{School of Physics and Astronomy, Sun Yat-sen University, No.2 Daxue Rd, 519082, Zhuhai China}
\affiliation[b]{State Key Laboratory of Physical Oceanography, Ocean University of China, Qingdao 266100, China}
\affiliation[c]{Joint Laboratory for Frontier Interdisciplinary Research in Ocean Science, Mathematics and Physics, Ocean University of China, Qingdao 266100, China}
\affiliation[d]{Department of Physics, University of Johannesburg, PO Box 524, Auckland Park 2006, South Africa}

\emailAdd{yanglli5@mail.sysu.edu.cn}
\emailAdd{cuiyd@mail.sysu.edu.cn}

\abstract{The physical origin of the diffuse neutrino background and its spectral break at $\sim$ 30 TeV remain a major puzzle in multi-messenger astrophysics. In this work, we demonstrate that this spectral feature is a natural consequence of AGN activity cycles and the resulting cosmic ray (CR) propagation. We present a unified model coupling the active and quiescent phases of AGNs, where CRs accelerated in the active core undergo subsequent diffusion and hadronic interactions in the host galaxy during the quiescent phase. The superposition of these distinct evolutionary phases yields dual spectral breaks, particularly the one at tens of TeV. Under realistic energetics, our model simultaneously accounts for the IceCube diffuse flux and fits the neutrino emissions of diverse sources, ranging from the blazar TXS 0506+056 to the Seyfert galaxies NGC 7469, CGCG 420-015, and the Circinus Galaxy. Our findings reveal that temporal variability is essential for deciphering the cosmic neutrino landscape and tracking high-energy CR escape.}

\begin{document}
\maketitle
\flushbottom

\section{Introduction}

In recent years, the IceCube Neutrino Observatory has established the existence of a diffuse neutrino background (DNB) spanning from TeV to PeV energies \citep{PhysRevLett.125.121104}. More recently, a combined fit (CF) analysis utilizing both track- and shower-like events has provided compelling evidence for a spectral break at approximately 30 TeV \citep{2gh9-d4q7}. The physical origin of this convex spectrum remains unclear and has attracted considerable attention.

As one of the most energetic classes of sources in the Universe, active galactic nuclei (AGNs) are prime accelerators of cosmic rays (CRs), potentially reaching EeV energies \citep{PhysRevLett.66.2697}. Their cores emit strong radiation fields from radio to soft $\gamma$-rays \citep{10.1111/j.1365-2966.2009.15007.x, PhysRevD.90.023007}, providing target photons for efficient neutrino production via photohadronic ($p\gamma$) interactions. Additionally, AGN host galaxies and clusters contain abundant gas on scales up to hundreds of kpc \citep{2008A&ARv..15...67M}, providing a site for proton-proton ($pp$) neutrino production. These properties make AGNs promising candidates for the dominant contributors to the observed DNB \citep{doi:10.1142}.

Various AGN models have been proposed to account for the DNB, including disk-corona \citep{PhysRevLett.125.011101}, radiatively inefficient accretion flows (RIAFs) \citep{2021NatCo..12.5615K}, Shakura-Sunyaev accretion disks \citep{2015JETP..120..541K}, inner jets \citep{PhysRevD.90.023007}, and CR reservoirs \citep{Hooper_2016, 2018NatPh..14..396F}. In the disk-corona scenario, protons accelerated by plasma turbulence generate a 10--100 TeV neutrino flux with a convex spectrum peaking at tens of TeV, consistent with IceCube measurements \citep{2026arXiv260220145M}. The substantial opacity of the dense core to GeV–TeV $\gamma$-rays naturally explains the isotropic diffuse $\gamma$-ray background (IGRB) observed by \textit{Fermi}-LAT \citep{Ackermann_2015}. When coupled with the RIAF or CR reservoir models, this scenario can account for the neutrino flux below 10 PeV \citep{2021NatCo..12.5615K, 2026arXiv260220145M}. Furthermore, the recent KM3NeT detection of KM3-230213A \citep{2025Natur.638..376K}, which is spatially associated with candidate blazars \citep{2026AJ....172..139A}, suggests that AGN inner jets could yield non-negligible contributions to the DNB.

% While these models offer valuable insights into specific energy regimes or source classes, a unified scenario capable of explaining the entire TeV–PeV DNB spectrum, particularly the dual spectral breaks, remains elusive. 
While these models offer valuable insights into specific energy regimes or source classes, a unified scenario capable of explaining the entire TeV–PeV DNB spectrum remains elusive. A common limitation of existing frameworks is the assumption of steady-state emission, which neglects the episodic nature and finite lifetimes of AGN activity. Because high-energy protons have cooling times far exceeding those of leptonic components, they can persist as relic CRs long after the central supermassive black hole (SMBH) engine shuts down. These relic CRs are subsequently confined within the magnetic fields of the host galaxy, interacting with the ambient interstellar gas to produce high-energy neutrinos via $pp$ collisions.

% While these theories offer various plausible mechanisms, the dominant origin of the DNB remains uncertain, and no single framework can fully reproduce the entire spectrum observed by IceCube. In addition, these models typically ignore the finite duration of AGN activity and the the subsequent evolution after the active phase ceases. Since protons cool far less efficiently than electrons, those accelerated during the active phase can survive long after activity ends. Such CRs may be confined within the host galaxy and later produce high-energy neutrinos through interactions with the ambient gas.

In this work, we propose a unified and time-dependent framework incorporating episodic activity of AGN (divided into active and quiescent phases) to investigate the origin of the DNB. We demonstrate that the observed spectrum can be interpreted as a cosmological superposition of AGNs at different evolutionary stages. During the active phase, neutrinos are primarily produced via $p\gamma$ interactions in the core and jet regions. In the quiescent phase, the relativistic protons that escaped from the jet and radio lobes are confined within the host galaxy, generating neutrinos through $pp$ interactions. We numerically model particle acceleration, cooling, and transport from the jet to the host galaxy, solving the diffusion equations self-consistently. Our model predicts a primary spectral break at several tens of TeV, which is shaped by the energy-dependent escape of CRs and the statistical age distribution of quiescent-phase AGNs. A secondary break at $\sim 10~{\rm PeV}$ naturally marks the transition of the dominant background component from quiescent-phase host-galaxy emission to active-phase jet photohadronic emission, consistent with recent IceCube and KM3NeT constraints. Furthermore, this framework successfully explains the multi-messenger signatures of individual candidate sources, including the active flaring blazar TXS 0506+056 \citep{doi:10.1126/science.aat1378} and the quiescent Seyfert galaxies NGC 7469, CGCG 420-015, and the Circinus Galaxy \citep{Abbasi_2026a, Abbasi_2026b}.

\section{Theoretical framework and model setup}
\label{model}
\subsection{AGN episodic activity and temporal simplification}

An AGN is powered by a central SMBH that accretes surrounding matter, releasing intense radiation and expelling relativistic particles. The accretion process can be triggered by galaxy mergers or internal galactic instabilities. Observations of local radio galaxies indicate that the active phase typically lasts $\sim10^{6}$--$10^{8}$ years, while the subsequent quiescent intervals are one to two orders of magnitude longer \citep{Turner_2015}. Furthermore, radio surveys show that only about 5--10\% of AGNs are radio‑loud (i.e., exhibit strong jets), while the majority are radio‑quiet \citep{Rafter_2009}, implying a very low duty cycle for jet activity.
% \citep{Jogee2006, 10.1111/j.1365-2966.2011.19669.x}

Based on this episodic nature, we define two distinct phases: the active phase, during which the SMBH is accreting and particle acceleration is ongoing, and the quiescent phase, which follows after the central engine shuts off. We adopt a representative duration for the AGN active phase as $T_{\rm act} = 1.0 \times 10^7~{\rm yr}$. Although the actual active lifetime of individual AGNs typically scales with the SMBH mass and the surrounding gas reservoir, adopting a constant $T_{\rm act}$ serves as a reasonable first-order approximation to minimize the number of free parameters. This simplification does not affect the robustness of our conclusions. Since the DNB is obtained by integrating the contributions from a vast population of AGNs over a wide redshift distribution ($z \sim 0 - 5$), the characteristic spectral break in the diffuse spectrum is predominantly determined by the average CR escape timescale in the host galaxies, rather than by the precise active lifetime of each source.

AGNs with lifetime $T\leq T_{\rm act}$ yr are accordingly classified as being in the active phase. The maximum lifetime $T_{\rm{max}}$ of AGN is assumed to be $1.0\times10^{10}$ yr, comparable to the Hubble time. During the active phase, the SMBH is fueled, and part of the accretion energy is channeled into accelerating particles, which form a relativistic jet that eventually interacts with the interstellar medium, creating a radio lobe that acts as a reservoir of nonthermal particles. During this phase, the AGN produces intense radiation, with neutrinos mainly generated through $p\gamma$ interactions. Light curve observations show evidence that AGN activity may have quasi-periodic modulation, with individual flares persisting for tens of days while the overall variability period extends for several years \citep{Zhang_2017a}. Thus, we further classify the active phase into a flare stage and a normal stage, with the flare stage having 100 times higher acceleration power and a 1\% duty cycle. These two parameters, together with the fraction of accretion energy that goes into particle acceleration, affect the overall normalization of the neutrino spectrum but do not influence its shape.

\begin{figure*}[!t]
    \centering
    \includegraphics[width=0.6\textwidth]{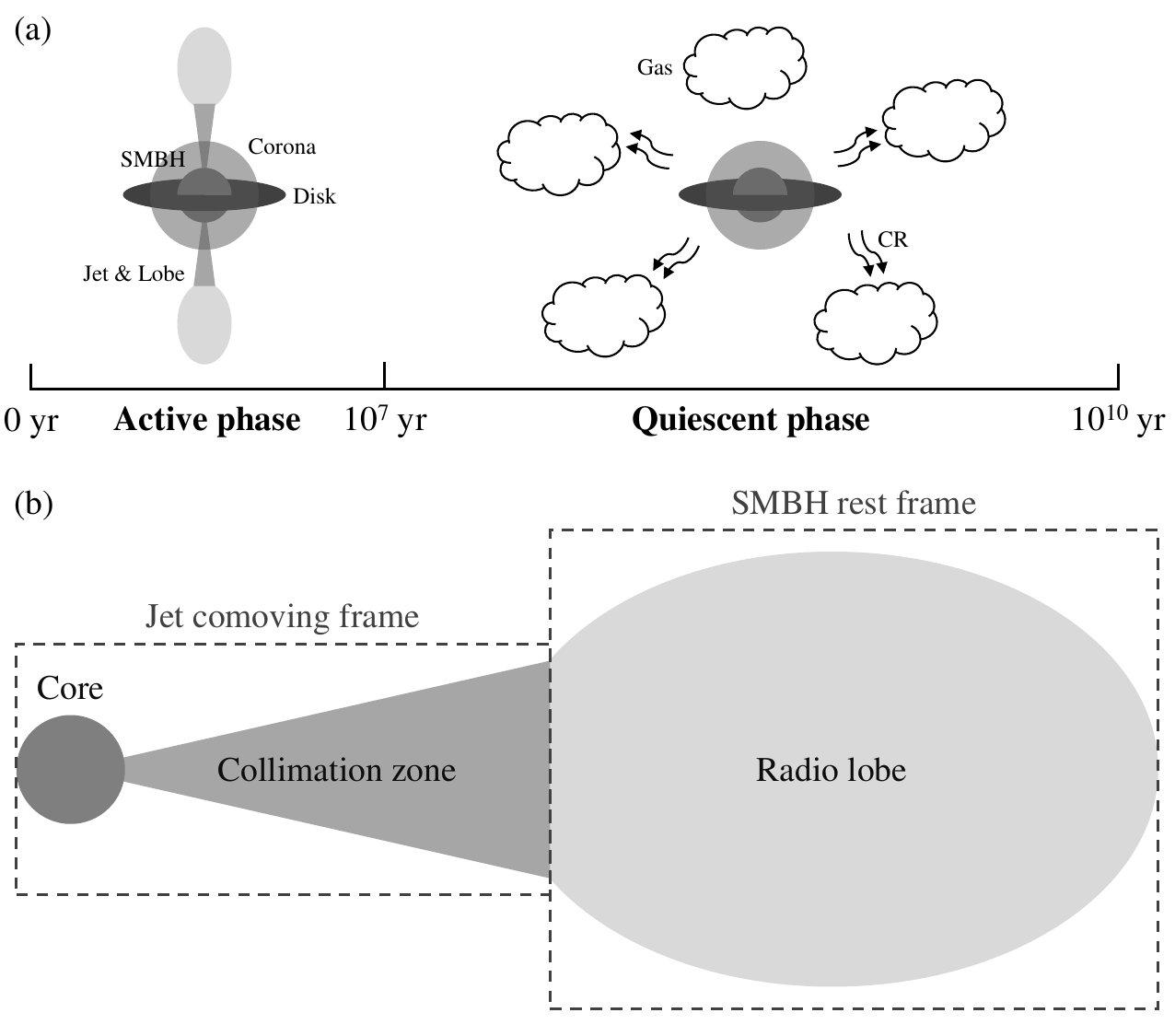}
    \caption{\label{fig:1}AGN variability and structure. Panel (a) illustrates the active and quiescent phases adopted in modeling AGN variability. Panel (b) shows the structure of an AGN during its active phase, highlighting the core, collimation zone, and radio lobe. The jet comoving frame is applied to the core region and collimation zone, while the SMBH rest frame is adopted for the lobe.}
\end{figure*}

During the quiescent phase, the power supply from the SMBH ceases, terminating particle injection and causing the lobe structure to gradually dissipate. High-energy protons previously confined within the lobe subsequently diffuse through the host galaxy and undergo $pp$ interactions with the ambient gas, thereby producing neutrinos. In certain scenarios, CRs may also escape into the host galaxy cluster; our model does not distinguish between these two prospects. The possibility of CRs escaping directly from the lobe into the intergalactic medium without interacting with the ambient gas is not considered. Figure~\ref{fig:1} (a) presents the schematic of our model for evaluating AGN variability.

The AGN structure also plays a critical role. During the active phase, an AGN typically comprises three components\textemdash the core (inner jet), the jet collimation zone, and the radio lobe, as shown in Fig.~\ref{fig:1} (b). Our model assumes primary particle acceleration occurs in the core region \citep{PhysRevD.90.023007, 2017SSRv..207....5R}, neglecting particle reacceleration in the collimation zone and lobe, as well as diffusive escape from all three regions. Owing to the relativistic effects of the jet, the core and collimation zone should be treated in the comoving frame, while for the lobe and host galaxy, the SMBH rest frame is used.
% this point will be elaborated in later subsections.

\subsection{Particle acceleration and propagation in the jets and radio lobe}

We describe the model for particle acceleration and propagation during the active phase of the AGN. Particles in astrophysical jets are expected to undergo diffusive shock acceleration (first-order Fermi acceleration; \citep{2007Ap&SS.309..119R, PhysRevD.90.023007}) or stochastic acceleration via turbulence (second-order Fermi acceleration;  \citep{Asano_2014}), though the relative dominance of these mechanisms remains a subject of active debate. Here, we assume that electrons and protons share a common acceleration mechanism, with the acceleration efficiency constrained by the empirical blazar sequence \citep{10.1046/j.1365-8711.1998.01828.x}. The corresponding acceleration and cooling timescales within the core region are calculated for both scenarios (see Fig.~S1 in the supplementary material). 

Under a typical first-order Fermi acceleration with a nominal efficiency parameter $\eta=0.05$, the comoving electron energy cutoff lies within the range of $10^{3}$ to $10^{4}$ GeV, whereas under second-order Fermi acceleration, it lies between 0.1 and 1 GeV. The corresponding proton cutoff energies span $10^{8}$--$10^{10}$ GeV and $10^{7}$--$10^{8}$ GeV for the first- and second-order scenarios, respectively. When transforming these values to the SMBH rest frame, they are scaled by the Doppler factor $\delta$ \citep{2007Ap&SS.311..231K}. 

The blazar sequence, which extends from radio to $\gamma$-ray bands, features a low-energy synchrotron peak at approximately 0.01 eV \citep{10.1046/j.1365-8711.1998.01828.x}.  This spectral peak corresponds to synchrotron emission from electrons with energies around $0.1~{\rm GeV}$, assuming a typical core magnetic field strength of $B \sim 1~{\rm G}$ \citep{PhysRevD.90.023007} and a representative Doppler factor of $\delta = 10$. This energetic constraint is naturally compatible with the second-order Fermi acceleration scenario, suggesting it as the more viable mechanism for reproducing the particle spectra required to reconcile the blazar sequence with the DNB. If an extremely low $\eta$ is adopted for first-order Fermi acceleration to match the electron cutoff at 0.1--1 GeV, the acceleration rate at TeV--PeV energies would be severely diminished. Consequently, the resulting maximum proton energies would fall short of the thresholds required to explain the high-energy DNB observed by IceCube. On these grounds, our framework adopts second-order Fermi acceleration as the primary acceleration mechanism. We refer to Supplementary material for detailed discussion.  

The temporal and spectral evolution of the particle distribution $n(E, t)$ within the core region under second-order Fermi acceleration is described by the following transport (Fokker-Planck) equation \citep{2017SSRv..207....5R, PhysRevLett.125.011101}:
\begin{align}
    \frac{\partial n(E,t)}{\partial t} =\;& \frac{\partial}{\partial E}\left[D_{E}(E)\frac{\partial n(E,t)}{\partial E} + \frac{E}{t_{\rm{cool}}(E)}n(E,t)\right] \nonumber\\
    \;& - \frac{n(E,t)}{t_{\rm{dyn}}} + Q(E,t).\label{eq-1}
\end{align}
Here, $n(E,t)$ is the particle number density per unit energy, $t_{\rm{cool}}(E)$ is the cooling timescale, and $t_{\rm{dyn}}$ is the characteristic dynamical timescale. The diffusion coefficient in momentum space for second-order Fermi acceleration is given by $D_{E}(E)=E^{2}/[(a+2)t_{\rm{acc-2}}(E)]$, where $a=v_{s}^{2}/v_{A}^{2}$ denotes the squared ratio of the shock velocity to the Alfvén velocity, and $t_{\rm{acc-2}}(E)$ represents the corresponding acceleration timescale. The injection term $Q(E,t)$ specifies the source feeding rate of particles into the acceleration zone. In our model, particles are injected and entrained by the relativistic jet with a bulk Lorentz factor $\Gamma$, taking the monoenergetic form \citep{2017SSRv..207....5R}
\begin{equation}\label{eq-2}
    Q(E,t)=Q_{0}(t)\delta(E-\Gamma m_{i}c^{2}),
\end{equation}
where $Q_{0}(t)$ is the normalization factor determined by the injected electron power $L_{e}$ (or proton power $L_{p}$) during the active flare stage, and $m_{i}$ is the rest mass of injected species. Both $L_{e}$ and $L_{p}$ are scaled with the AGN bolometric luminosity $L_{\rm{bol}}$. Since $L_{\rm{bol}}$ is generally challenging to measure directly, we utilize the empirical correlation with the X-ray luminosity $L_{X}$ in the $2$--$10~{\rm keV}$ band \citep{Hopkins_2007} to express $L_{e}$ and $L_{p}$ as functions of $L_{X}$ \citep{2017SSRv..207....5R, PhysRevLett.125.011101}:
\begin{align}\label{eq-3}
    L_{e}=\;&\frac{1}{2}q_{e}q_{\rm{rel}}q_{\rm{jet}}\dot{M}c^{2}\nonumber\\
    =\;&\frac{1}{2}q_{e}q_{\rm{rel}}q_{\rm{jet}}\frac{L_{\rm{bol}}}{\eta_{\rm{rad}}}\nonumber\\
    =\;&\frac{1}{2}q_{e}q_{\rm{rel}}q_{\rm{jet}}\frac{k_{\rm{bol/X}}L_{\rm{X}}}{\eta_{\rm{rad}}},\\
    \label{eq-4}L_{p}=\;&\frac{1}{2}q_{p}q_{\rm{rel}}q_{\rm{jet}}\frac{k_{\rm{bol/X}}L_{\rm{X}}}{\eta_{\rm{rad}}}.
\end{align}

Here, $q_{e}$ and $q_{p}=1-q_{e}$ represent the respective power fractions partitioned into relativistic electrons and protons. 
The parameter $q_{\rm{rel}}<1$ denotes the efficiency of converting jet bulk kinetic energy into non-thermal particles, while $q_{\rm{jet}}<1$ is the fraction of the total accretion power channeled into the jet. 
The radiative efficiency of the accretion disk is denoted by $\eta_{\rm rad}$, and $k_{\rm bol/X}$ represents the bolometric correction factor. We adopt the empirical scaling relation $k_{\rm{bol/X}}=15[L_{X}/(1.0\times10^{42}\,\rm{erg}\,\rm{s^{-1}})]^{0.35}$. To break degeneracy and constrain the product $q_{\rm{rel}}q_{\rm{jet}}$, we require that the electron power $L_{e}$, derived from Eq.~(\ref{eq-3}), yields synchrotron and inverse-Compton (IC) radiation that matches the observed blazar sequence. The value of $L_{p}$ is uniquely determined by prescribing the baryon loading factor $L_{p}/L_{e}$. 
Beyond its tight correlation with $L_{\rm bol}$, X-ray luminosity $L_X$ is selected as the primary proxy for the particle energy budget because X-rays are ubiquitously detected across almost all AGN classes, unlike radio or $\gamma$-ray emissions which are highly state- or subclass-dependent.

We solve Eq.~(\ref{eq-1}) numerically using the Chang-Cooper finite-difference scheme \citep{park1996fokker} to track the time-dependent nonthermal spectra of electrons and protons in the core region, setting $t = 0$ at the onset of the active phase. A similar transport scheme is applied sequentially to the downstream collimation zone and the radio lobe. For these outer regions, momentum diffusion is omitted, and the injection term $Q(E, t)$ is set equal to the particle flux dynamically escaping from the immediate upstream boundary (e.g., from the core to the collimation zone), preserving particle number. Specifically, these upstream escape spectra are evaluated by solving the transport equation Eq.~(\ref{eq-1}) up to $t = t_{\rm dyn}$ without the loss term $-n(E, t)/t_{\rm dyn}$ to isolate the spectral reshaping caused purely by acceleration and cooling. Crucially, we assume that all CRs remaining in the radio lobe at the end of the active cycle enter the quiescent phase without undergoing dynamical escape during the active phase. After resolving the spatial and spectral evolution of the electrons, we utilize the radiation modeling package \texttt{GAMERA} \citep{2015ICRC...34..917H} to compute the multiwavelength photon spectra. These local radiation fields serve as targets for calculating proton energy losses via $p\gamma$ interactions. The resulting neutrino production is calculated using the Sim-B code \citep{Hummer_2010}. All calculations are performed in the jet comoving frame for the core and collimation zone, while the radio lobe and host galaxy environments are treated in the SMBH rest frame, as illustrated in Fig.~\ref{fig:1} (b).

\subsection{Cosmic ray diffusion and hadronic interactions in host galaxies}

During the quiescent phase of the AGN, CRs that were previously confined within the radio lobe during the active phase escape into the host galaxy, where they diffuse and undergo hadronic interactions. To simulate this long-term propagation, we employ the spatially-averaged leaky-box approximation \citep{refId4}, which serves as a computationally efficient and reliable representation of the standard spatial diffusion equation under galactic boundary conditions \citep{Cowsik_2025}. The temporal and spectral evolution of the CR proton distribution is governed by
\begin{equation}\label{eq-5}
	\frac{\partial n(E,t)}{\partial t}=\frac{\partial}{\partial E}\left[\frac{E}{t_{\rm{cool}}(E)}n(E,t)\right]-\frac{n(E,t)}{t_{\rm{esc}}(E)},
\end{equation}
where $t_{\rm cool}(E)$ represents the total cooling timescale dominated by $pp$ collisions, and $t_{\rm esc}(E)$ is the energy-dependent diffusive escape timescale, which is parameterized as $t_{\rm esc}(E) \approx H_g^2 / [\pi^2 D_S(E)]$, where $H_g$ denotes the characteristic scale height of the host galaxy, and $D_S(E)$ is the energy-dependent diffusion coefficient.

We model the transition from the active to the quiescent phase as a sudden injection of particles into the host galaxy. Specifically, at $t = T_{\rm act} = 1.0 \times 10^7~{\rm yr}$ (the duration of the active phase), the central engine shuts off, and the magnetically confined CRs in the radio lobe are released into the galactic environment. This scenario is physically justified since the typical diffusion timescale for PeV CRs to escape the relic lobe is on the order of several million years \citep{2018NatPh..14..396F}, which is comparable to the active lifetime $T_{\rm act}$. Consequently, we take the final-state CR spectrum in the lobe obtained from Eq.~(\ref{eq-1}) and evaluated at $t = T_{\rm act}$ as the initial condition ($t_{\rm qui} = 0$, with $t_{\rm qui} \equiv t - T_{\rm act}$ the time elapsed since the central engine shutdown) to solve Eq.~(\ref{eq-5}) during the quiescent phase. 

Equation~(\ref{eq-5}) is solved numerically using the Chang-Cooper scheme to yield the time-dependent CR spectrum $n(E, t_{\rm qui})$ over the quiescent duration $t_{\rm qui}$. Utilizing these evolving CR spectra, we calculate the production of secondary pions and the resulting neutrino spectra via hadronic $pp$ interactions with the ambient cold interstellar gas of density $n_p$. The secondary particle yields and the final neutrino spectra are computed using the method in \cite{PhysRevD.74.034018}.
% Our model assumes that all CRs leave the radio lobe at $t=1.0\times10^{7}$ yr, equal to the lifetime of the active phase, and subsequently enter the host galaxy. This assumption is plausible, as the typical timescale for PeV CRs to escape the lobe via diffusion is several Myr \citep{2018NatPh..14..396F}, which is comparable to the lifetime of the active phase. Taking the final-state CR spectrum in the lobe (obtained by solving Eq.~(\ref{eq-1}) at $t=1.0\times10^{7}$ yr) as the initial condition, we numerically solve Eq.~(\ref{eq-5}) using the Chang-Cooper method to obtain the time-dependent CR spectrum during the quiescent phase. We then incorporate the $pp$ interaction model from \cite{PhysRevD.74.034018} to derive the corresponding time-dependent neutrino spectrum. 

\section{Results and analysis}

\subsection{The cosmic diffuse neutrino background and spectral breaks}

Based on the theoretical framework of AGN variability and spectral evolution detailed in Section \ref{model}, we now calculate the cumulative DNB and investigate whether its characteristic spectral features can be naturally explained by episodic AGN activity. 
Since the diffusion coefficient increases with $E$, higher-energy CRs have shorter $t_{\rm{esc}}(E)$, leading to a progressive softening of the CR spectrum within the host galaxy during the quiescent phase compared to the initial injection spectrum during the active phase. We therefore expect the sub-PeV component of the DNB to be dominated by the diffuse emission from AGNs in their quiescent phase, whereas the higher-energy background is primarily shaped by active sources. The total DNB at Earth is evaluated by performing a cosmological volume integral over redshift $z$, AGN active lifetime $T$, and X-ray luminosity $L_X$ \citep{PhysRevLett.125.011101, 2021NatCo..12.5615K}: 
\begin{align}\label{eq-6}
\Phi_{\nu+\bar{\nu}}(E_{\nu})=\;&\frac{c}{4\pi H_{0}}\int_{0}^{z_{\rm max}} \frac{\mathrm{d}z}{\sqrt{(1+z)^{3}\Omega_{m}+\Omega_{\Lambda}}}\nonumber\\\nonumber\;&\times\int_{0}^{T_{\rm max}} \frac{\mathrm{d}T}{T_{\rm max}} \int_{L_{X, \rm min}}^{L_{X, \rm max}} \mathrm{d}L_X \frac{\mathrm{d}\rho_X}{\mathrm{d}L_X}(z)\\\;&\times\frac{L_{E_\nu'}(E_\nu, z, T, L_X)}{E_\nu'},
	% \Phi_{\nu+\bar{\nu}}(E_{\nu})&=&\frac{c}{4\pi H_{0}}\int dz\frac{1}{\sqrt{(1+z)^{3}\Omega_{m}+\Omega_{\Lambda}}}\nonumber\\&&\times\iint dTdL_{X}\frac{1}{T_{\rm{max}}}\frac{d\rho_{X}}{dL_{X}}(z)\nonumber\\&&\times\frac{L_{E'_{\nu}}}{E'_{\nu}}(E_{\nu},z,T,L_{X}),
\end{align}
where $H_{0}=67.8$ km s$^{-1}$ Mpc$^{-1}$ represents the Hubble constant \citep{refId5}, with cosmological parameters $\Omega_{m}=0.3$ and $\Omega_{\Lambda}=0.7$. We assume the AGN lifetime $T$ is uniformly distributed over the interval $[0, T_{\rm max}]$ with $T_{\rm max} = 1.0 \times 10^{10}~{\rm yr}$, which is comparable to the Hubble time. The term $d\rho_{X}/dL_{X}$ stands for the X-ray luminosity function of AGNs per unit comoving volume per unit luminosity interval, for which we adopt the widely used model and parameters from \cite{Ueda_2014}. The quantities $E_\nu$ and $E_\nu' = (1+z)E_\nu$ represent the observed neutrino energy at Earth and the intrinsic neutrino energy at redshift $z$, respectively. Finally, $L_{E_\nu'}$ is the intrinsic neutrino luminosity per unit energy of an individual source.

\begin{figure*}[!t]
    \centering
    \includegraphics[width=0.75\textwidth]{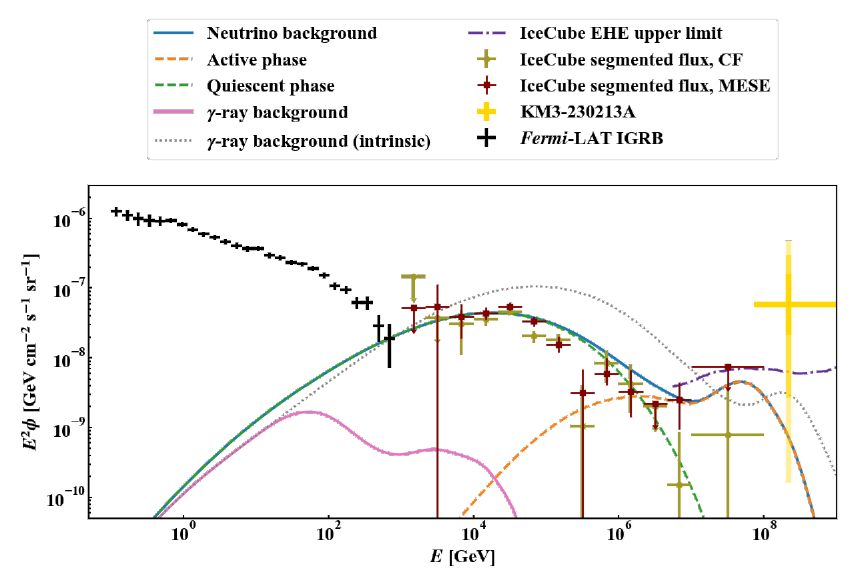}
    \caption{\label{fig:2}Model-predicted diffuse neutrino flux per neutrino flavor and the accompanying diffuse $\gamma$-ray flux, compared with IceCube, KM3NeT and \textit{Fermi}-LAT data. The orange and green dashed lines show contributions from AGNs in the active and quiescent phases, respectively; the blue solid line is their sum. The purple dash-dotted line denotes the IceCube extremely-high-energy (EHE) upper limit \citep{PhysRevD.98.062003}. Points with error bars in dark khaki and crimson indicate the IceCube segmented flux derived from the CF approach and the MESE data \citep{2gh9-d4q7}. The gold cross stands for the flux inferred from KM3-230213A \citep{2025Natur.638..376K}, with the horizontal span illustrating the central 90\% neutrino energy range and vertical bars representing the 1$\sigma$, 2$\sigma$, and 3$\sigma$ Feldman-Cousins confidence intervals on the flux estimate. The pink solid and gray dotted lines are the accompanying diffuse $\gamma$-ray flux at Earth and its intrinsic value (without EBL absorption). The \textit{Fermi} IGRB \citep{Ackermann_2015} is displayed by black crosses.}
\end{figure*}

By calculating the time-dependent neutrino spectrum $L_{E_\nu'}$ from our physical model, we apply Eq.~\eqref{eq-6} to obtain the DNB. We present the resulting diffuse flux per neutrino flavor in Fig.~\ref{fig:2} and compare it with observational data. The benchmark model parameters are summarized in Table~\ref{tab:1}. For the host galaxy environment, we adopt a representative cold interstellar gas density of $n_p = 1~{\rm cm^{-3}}$, consistent with the average value observed in the Galactic plane \citep{10.1093/mnras/stx1580}. Integration of redshifts takes place from $z = 0$ to $z = 5$. Further details are given in the supplementary material.

As shown in Fig.~\ref{fig:2}, the observed DNB can be successfully reproduced as a superposition of neutrino emissions from AGNs at different evolutionary stages. Quiescent-phase AGNs dominate the diffuse flux at energies below $1~{\rm PeV}$, while active-phase AGNs account for the higher-energy component. The neutrino spectrum from active AGNs exhibits two distinct peaks. The high-energy peak at $\sim 50~{\rm PeV}$ is dominated by $p\gamma$ interactions in the highly boosted relativistic jets (principally from blazars), whereas the lower-energy component extending down to the TeV band is dominated by emissions from the radio lobes.

Importantly, our model predicts a convex spectrum with a distinct spectral break at several tens of TeV, which is in agreement with the latest IceCube CF and medium energy starting events (MESE) data. This spectral break is a natural consequence of the competition between the temporal evolution of the CR reservoir and the population statistics of quiescent AGNs. According to Eq.~(\ref{eq-5}), the CR intensity in the host galaxy decays approximately as $\exp[-t_{\rm qui}/t_{\rm esc}(E)]$. In the early quiescent phase (i.e., $t_{\rm qui}<t_{\rm{esc}}(E)$), the CR spectrum is hard and its intensity varies rapidly with time. At later times (i.e., $t_{\rm qui}>t_{\rm{esc}}(E)$), the CR spectrum softens due to the diffusive escape of high-energy protons, while these aged sources are statistically far more numerous. Consequently, the trade-off between the small number of quiescent AGNs with hard spectra and the abundant ones with softer spectra gives rise to the spectral break. 

The precise energy of this break is determined by the diffusive escape timescale $t_{\rm{esc}}(E)$. To reproduce the break at $\sim30~{\rm TeV}$, our model requires $t_{\rm esc} \gtrsim 10^7~{\rm yr}$ for PeV protons. This timescale is longer than the typical galactic escape timescales of a few million years estimated previously \citep{Hooper_2016}, suggesting that AGN host galaxies are highly efficient CR reservoirs.

\renewcommand\arraystretch{1.25}
\begin{table*}[!t]
  \centering
  \caption{\label{tab:1}Primary parameters adopted in DNB modeling}
  \begin{tabular}{ccc}
  \hline
    Parameter & Adopted value & Typical value \\ \hline
    $q_{\rm{rel}}q_{\rm{jet}}$ & 0.5\% & 0.1\%--1\% \citep{refId6} \\
    $L_{p}/L_{e}$ & 0.015 & $<1$ for leptonic model \citep{2017SSRv..207....5R} \\
    $\eta_{\rm{rad}}$ & 0.1 & 0.1 \citep{PhysRevLett.125.011101} \\
    $H_{g}$ [kpc] & 20 & 1-100 \citep{10.3389/fspas.2021.694554} \\
    $D_{S}(E)$ [cm$^{2}$ s$^{-1}$] & $1.0\times10^{28}\times\newline(E/4\,\mathrm{GeV})^{0.29}$ & $6.5\times10^{28}\times\newline(E/4\,\mathrm{GeV})^{0.29}$ \citep{PhysRevD.103.115010} \\
    \hline
  \end{tabular}
\end{table*}
\renewcommand\arraystretch{1.0}

Furthermore, a secondary spectral break is visible at $\sim10$ PeV in Fig.~\ref{fig:2}, which is consistent with the CF data from IceCube \citep{2gh9-d4q7} and the flux inferred from KM3-230213A \citep{2025Natur.638..376K}. This feature marks the transition of the dominant emission component contributing to the DNB. Below $10~{\rm PeV}$, the DNB is dominated by the combination of radio lobe emission from active sources and the diffuse $pp$ interactions in the host galaxies of early-stage quiescent AGNs. Above $10~{\rm PeV}$, however, the intense radiation fields in the jet of active AGNs (especially blazars) trigger highly efficient $p\gamma$ interactions, resulting in a harder spectral component that dominates at the highest energies.

% As previously discussed, the diffuse neutrino flux from active AGNs comprises two major components, with the high-energy part originating from jet emission. Particles in the jet\textemdash especially in the inner jet\textemdash are more energetic than those in the radio lobe, and the radiation fields in the vicinity of the SMBH are more intense and complex \citep{10.1111/j.1365-2966.2009.15007.x}. 

\begin{figure*}[!t]
    \centering
    \includegraphics[width=0.6\textwidth]{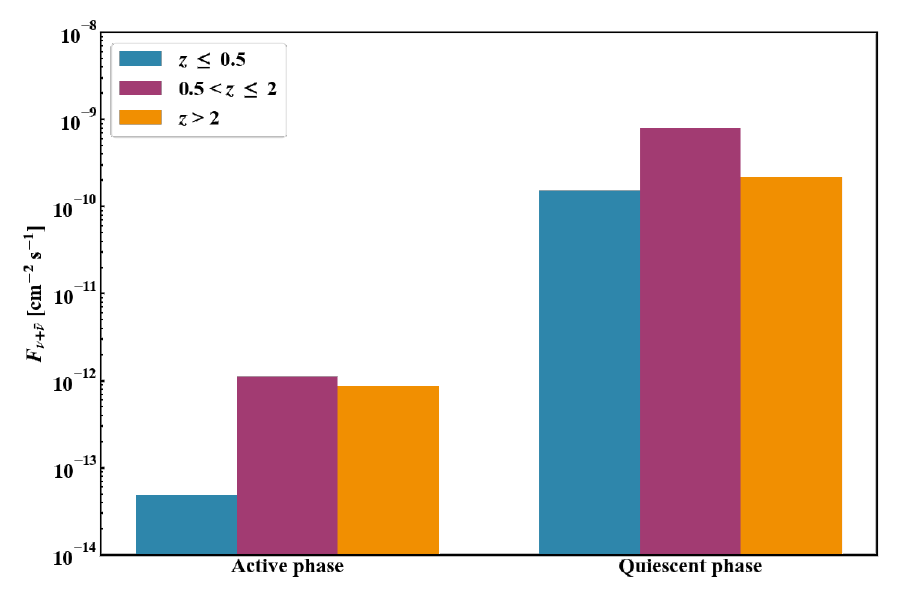}
    \caption{\label{fig:3}Histogram of the DNB composition. Blue, purple, and orange bars represent the all-flavor neutrino count per unit area per unit time from AGNs at low-, intermediate-, and high-redshifts, respectively, integrated from 1 TeV to 1 EeV. The two sets of bars correspond to contributions from the active and quiescent phases.}
\end{figure*}

To investigate the cosmological origin of the DNB, we divide the AGN population into three redshift intervals: low ($z \le 0.5$), intermediate ($0.5 < z \le 2$), and high ($z > 2$). The relative contributions of these bins to the total DNB (integrated from $1~{\rm TeV}$ to $1~{\rm EeV}$) are shown in Fig.~\ref{fig:3}. The primary component of the DNB originates from intermediate- and high-redshifts AGNs, whereas low-redshift AGNs contribute only about 13\%. Furthermore, because the CR confinement timescale in the host galaxy ($t_{\rm esc} \sim 10^7 - 10^9~{\rm yr}$) vastly exceeds the active lifetime of the jet and lobe ($T_{\rm act} = 1.0 \times 10^7~{\rm yr}$), our scenario predicts that the neutrino sky is overwhelmingly dominated by quiescent-phase AGNs, while active-phase AGNs contribute less than $1\%$ of the cumulative flux. 
% This result is consistent with recent statistical stacking analyses of 10-year IceCube data \citep{2026arXiv260202390J}.

As summarized in Table~\ref{tab:1}, our model adopts a diffusion coefficient $D_{S}(E)=1.0\times10^{28}(E/4\,\mathrm{GeV})^{0.29}$ cm$^{2}$ s$^{-1}$, which is smaller than the Galactic value \citep{PhysRevD.103.115010} and comparable to that of the radio lobe \citep{Mathews_2007}. This value leads to a longer $t_{\rm{esc}}(E)$ than previously assumed and is physically plausible. Relic radio lobes are observed to persist for $\sim10^{6}$--$10^{7}$ years after the central engine shuts off \citep{refId0}\footnote{The relic phase in \cite{refId0} refers to the period when the central engine has ceased but radio emission still persists; it is not equivalent to the quiescent phase defined in this work.}. The residual turbulent magnetic field in these relic structures can therefore hinder CR escape, rendering the host galaxy an exceptionally efficient reservoir during the post-active phase. This picture is further supported by the fact that high-redshift galaxies generally exhibit stronger magnetic fields \citep{Kronberg_2008}. Additionally, the inferred low baryon loading factor ($L_p / L_e = 0.015$) indicates that the non-thermal emission from the majority of AGN jets is dominated by leptonic processes. 
% \citep{10.1111/j.1365-2966.2006.10986.x, Bottcher_2013} 
% This conclusion naturally explains the relatively low detection rate of point sources by IceCube.

Finally, because neutrino production via charged pion decay is inevitably accompanied by neutral pion decay into high-energy $\gamma$-rays, we calculate the accompanying diffuse $\gamma$-ray background using a method analogous to that employed for the neutrino background. The resulting $\gamma$-ray flux is shown in Fig.~\ref{fig:2}. Since the dominant contributors to the DNB reside at intermediate to high redshifts $z>0.5$, the co-produced $\gamma$-rays suffer severe attenuation via pair production on the extragalactic background light (EBL) during propagation \citep{Finke_2022}. As a result, the predicted diffuse $\gamma$-ray flux at Earth drops steeply in the TeV band, remaining well below the IGRB measured by \textit{Fermi}-LAT \citep{Ackermann_2015}, thereby ensuring self-consistency with current $\gamma$-ray observations. 

\subsection{Multi-messenger emissions from individual active and quiescent sources}

The DNB is fundamentally a superposition of emissions from individual AGNs. Observations of individual neutrino sources are vital for identifying the origins of both ultra-high-energy cosmic rays (UHECRs) and neutrinos. Recent IceCube searches have identified several compelling point-source candidates, including the flaring blazar TXS 0506+056, and nearby Seyfert galaxies such as NGC 1068, NGC 4151, NGC 7469, CGCG 420-015, and the Circinus Galaxy. Various models\textemdash such as the disk-corona scenario \citep{Carpio_2026} and the outflow-cloud interaction model \citep{2025arXiv251106707M}\textemdash have been proposed to explain these emissions. 
% For NGC 1068 and NGC 4151, however, stringent $\gamma$-ray upper limits set by \textit{Fermi}-LAT \citep{Ajello_2023, Murase_2024} and MAGIC \citep{Acciari_2019, refId7} in the 100 GeV--10 TeV range make these sources difficult to reproduce with our model. They may represent special cases requiring alternative scenarios. A more detailed discussion is provided in the supplementary material.

Among them, NGC 1068 and NGC 4151 are well-known to be highly obscured, Compton-thick Seyfert II galaxies. Their neutrino emissions are widely believed to originate from the innermost, optically thick corona region where high-energy $\gamma$-rays are heavily absorbed [e.g., \citep{Ajello_2023, Murase_2024}]. In contrast, our transient diffusion model is naturally applicable to Seyfert galaxies during their quiescent phases, where the host galaxy as CR reservoir dominates the diffuse emissions. This suggests a dual-population origin for Seyfert galaxy neutrinos, so that coronal emission in actively accreting, heavily obscured cores, and diffuse host-galaxy emission in post-active sources. Consequently, we focus our individual source modeling on TXS 0506+056 (active phase), and NGC 7469, CGCG 420-015, and the Circinus Galaxy (quiescent phase).

\begin{figure*}[!t]
    \centering
    \includegraphics[width=0.9\textwidth]{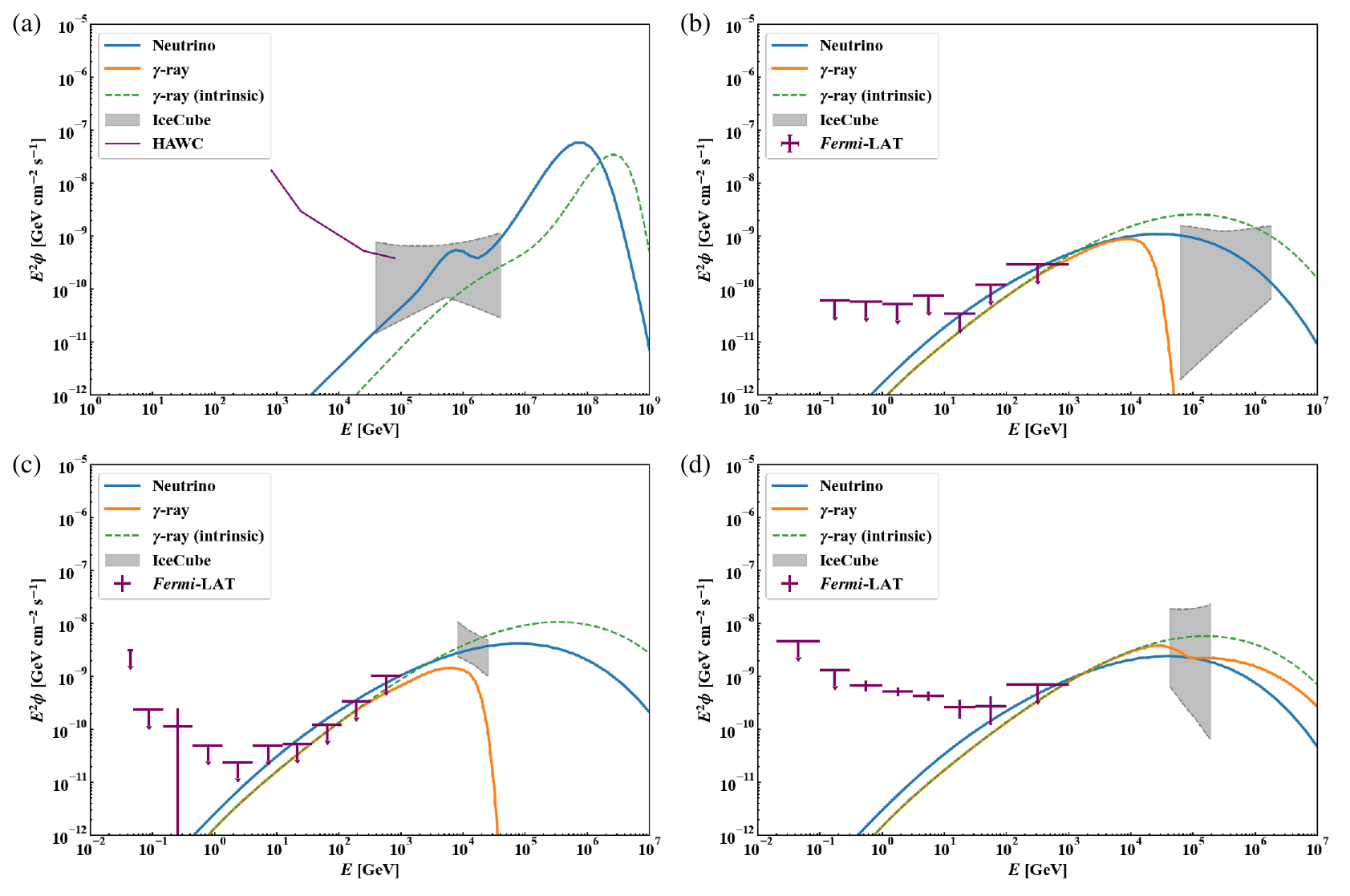}
    \caption{\label{fig:4}Interpretation of individual sources. Panel (a) presents the expected neutrino flux per neutrino flavor of TXS 0506+056 and the accompanying $\gamma$-ray spectrum, compared with the IceCube butterfly plot \citep{doi:10.1126/science.abg3395} and the HAWC upper limit \citep{doi:10.1126/science.aat1378}. The orange solid line represents the $\gamma$-ray flux at Earth, which is at a level of $10^{-16}$ GeV cm$^{-2}$ s$^{-1}$ and is too low to be visible on the figure. Panel (b) gives the expected neutrino and $\gamma$-ray fluxes from NGC 7469, alongside the IceCube \citep{Abbasi_2026a} and \textit{Fermi}-LAT \citep{Yang_2025} data. Panel (c) displays the corresponding fluxes from CGCG 420-015, compared with IceCube \citep{Abbasi_2026a} and \textit{Fermi}-LAT \citep{2025arXiv251106707M} observations. Panel (d) shows the results for the Circinus Galaxy, together with IceCube \citep{Abbasi_2026b} and \textit{Fermi}-LAT \citep{Murase_2024} data.}
\end{figure*}

The modeling results for these individual sources are presented in Fig.~\ref{fig:4}, and the key physical parameters including X-ray luminosity $L_{X}$ and luminosity distance $d_{L}$ are summarized in Table~\ref{tab:2}. The values for $\eta_{\rm{rad}}$, $H_{g}$ and $D_{S}(E)$ remain identical to those used in the DNB calculation. For the flaring blazar TXS 0506+056, we assume it is in the active phase ($T = 3 \times 10^6~{\rm yr}$). As shown in Fig.~\ref{fig:4} (a), the predicted neutrino spectrum exhibits a prominent peak at several hundred TeV. This peak originates from the decay of neutrons, which are produced via hyperon resonances generated in interactions between high-energy protons and the electron synchrotron/IC radiation field near the energy threshold. Because neutrons are charge-neutral, they escape the intense magnetic fields of the acceleration zone before decaying into protons, electrons, and electron antineutrinos ($n \to p + e^- + \bar{\nu}_e$) \citep{Hummer_2010}. The peak energy of this neutron-decay component is in good agreement with the high-energy neutrino event IceCube-170922A associated with TXS 0506+056 \citep{doi:10.1126/science.aat1378}, assuming that the produced $\bar{\nu}_e$ undergoes standard three-flavor oscillation to $\bar{\nu}_\mu$ during propagation. The absence of the higher-energy pion-decay peak in current observations can be attributed to the limited exposure and lower sensitivity of IceCube at EeV energies.
As shown in Table~\ref{tab:2}, the required baryon loading factor $L_{p}/L_{e}$ for TXS 0506+056 is more than an order of magnitude higher than the value used for the DNB and Seyfert galaxies. Such an elevated baryon loading in flaring blazars is physically consistent with recent multi-messenger leptohadronic modeling of individual active states \citep{refId8}.

\renewcommand\arraystretch{1.25}
\begin{table*}[!t]
  \centering
  \caption{\label{tab:2}Primary parameters adopted in individual source modeling}
  \begin{tabular}{ccccc}
  \hline
     & TXS 0506+056 & NGC 7469 & CGCG 420-015 & Circinus Galaxy\\ \hline
    $L_{X}$ [erg s$^{-1}$] & $1\times10^{44}$ \citep{Fiorillo_2025} &  $1.3\times10^{43}$ \citep{Testagrossa_2026} & $1\times10^{44}$ \citep{Testagrossa_2026} & $3.1\times10^{42}$ \citep{Carpio_2026} \\
    $d_{L}$ [Mpc] & 1774 \citep{Fiorillo_2025} & 71 \citep{Testagrossa_2026} & 134 \citep{Testagrossa_2026} & 4.1 \citep{Carpio_2026} \\
    $T$ [yr] & $3\times10^{6}$ & $3\times10^{7}$ & $2\times10^{7}$ & $2\times10^{7}$ \\
    $n_p$ [cm$^{-3}$] & / & 1 & 0.3 & 0.05 \\
    $q_{\rm{rel}}q_{\rm{jet}}$ & 0.75\% & 0.5\% & 0.5\% & 0.5\% \\
    $L_{p}/L_{e}$ & 0.5 & 0.015 & 0.015 & 0.015 \\
    \hline
  \end{tabular}
\end{table*}
\renewcommand\arraystretch{1.0}

Figure~\ref{fig:4} (b)-(d) present the modeling results for NGC 7469, CGCG 420-015, and the Circinus Galaxy, respectively. These three sources are Seyfert galaxies whose current states are consistent with the quiescent phase in our model, with ages since the shutdown of active phase injection set to $T \sim (2-3) \times 10^7~{\rm yr}$. As illustrated, our model successfully reproduces the observed neutrino fluxes, matching the IceCube butterfly data. Crucially, the accompanying hadronic $\gamma$-ray fluxes at Earth remain well below the upper limits established by \textit{Fermi}-LAT. Since our model predicts detectable $\gamma$-ray emissions in the $1$--$10~{\rm TeV}$ range for NGC 7469 and CGCG 420-015, and extending up to PeV energies for the Circinus Galaxy, future observations by very-high-energy (VHE) and ultra-high-energy (UHE) $\gamma$-ray observatories—such as MAGIC, VERITAS, H.E.S.S., HAWC, and LHAASO—will serve as a critical diagnostic to validate this quiescent-phase reservoir scenario.

\section{Discussion and conclusions}

The recent discovery of a DNB with convex spectrum detected by IceCube from TeV to PeV challenges conventional steady-state astrophysical scenarios and hints at novel particle acceleration or transport physics. In this work, we have developed a unified, time-dependent framework incorporating AGN episodic activity (active and quiescent phases) to simulate particle acceleration and propagation in the jet, collimation zone, and radio lobe, followed by CR diffusion and hadronic interactions in the host galaxy. Our model successfully reproduces the observed DNB spectrum, naturally explaining the two prominent spectral breaks. The break at several tens of TeV arises from the temporal evolution of the CR reservoir and the population statistics of quiescent-phase AGNs, while the secondary break at $\sim 10~{\rm PeV}$ marks the transition of the dominant background component from radio-lobe and host-galaxy emission to jet-dominated $p\gamma$ processes. We show that the DNB can be interpreted as a cosmological superposition of AGNs at different evolutionary stages. Furthermore, this framework successfully models the multi-messenger emissions from representative individual sources, namely the active flaring blazar TXS 0506+056 and the quiescent-phase Seyfert galaxies NGC 7469, CGCG 420-015, and the Circinus Galaxy.

% Overall, these results provide a novel perspective that the DNB is the superposition of neutrino emission from AGNs at different variation phases, and all its spectral features are consequences of AGN variability. Additionally, our model can be applied to interpret the neutrino fluxes from four representative sources: TXS 0506+056, NGC 7469, CGCG 420-015, and the Circinus Galaxy. 

A key finding of our model is the requirement of a suppressed diffusion coefficient in the quiescent host galaxies, $D_S(E) = 1.0 \times 10^{28} (E / 4~{\rm GeV})^{0.29}~{\rm cm^2~s^{-1}}$, which is smaller than the nominal Galactic value. This implies that AGN host galaxies must act as highly efficient CR reservoirs, retaining PeV protons for timescales exceeding $10^7~{\rm yr}$. As discussed in the Supplemental Material, this scenario is physically supported by several independent observational and theoretical evidence. Relic radio lobes are observed to persist for $10^6$--$10^7~{\rm yr}$ after the shutoff of the central engine \citep{refId0}, and the decaying magnetic turbulence within these relic structures can strongly inhibit CR diffusion. This confinement is further enhanced by the stronger magnetic fields typical of high-redshift galaxies \citep{Kronberg_2008}, the presence of "fossil" AGN ionization zones \citep{10.1093/mnras/stt1150}, and diffusion suppression in active star-forming or starburst regions \citep{Semenov_2021}.

The relatively low baryon loading factor ($L_p/L_e = 0.015$) inferred for the bulk of the AGN population in our model points to a leptonic-dominated emission scenario for typical AGNs, which is consistent with empirical constraints derived from cluster environments \citep{Shi_2022}. This follows naturally from the baseline assumption that the electromagnetic radiation defining the blazar sequence originates from leptonic origin. Crucially, however, the low baryon loading does not decouple the physical link between high-energy neutrinos and UHECRs. While a low $L_p/L_e$ for the general population constrains the average CR energy budget, local target densities (both gas and photons) in individual sources remain sufficient to power detectable neutrino fluxes. Indeed, our modeling of representative sources shows that the adopted gas densities and radiation fields are physically reasonable for both $pp$ and $p\gamma$ scenarios. The resulting neutrino fluxes agree with IceCube observations and with predictions from the leptohadronic calculations \cite{refId8}. This self-consistency demonstrates that the CR densities employed in our framework are compatible with those in conventional scenarios.

% The low baryon loading factor ($L_p/L_e = 0.015$) inferred for the bulk of the AGN population in our model favors leptonic-dominated emission models for typical AGNs, which is consistent with observational constraints from cluster environments \citep{Shi_2022}. This low hadronic component suggests that the sub-PeV neutrino flux modeled here constitutes a distinct physical component produced by moderate-energy protons ($\lesssim 100~{\rm PeV}$) confined within host galaxies, thus decoupling it from the primary acceleration sites of UHECRs. Nonetheless, this does not exclude a physical link between neutrinos and UHECRs during transient flaring states, as demonstrated by the elevated baryon loading ($L_p/L_e = 0.5$) required to fit the flaring blazar TXS 0506+056. 

Our model offers a self-consistent, unified explanation for the entire TeV--PeV neutrino spectrum. While alternative scenarios, such as the AGN disk-corona model, successfully explain the $10$--$100~{\rm TeV}$ neutrino flux and the associated MeV $\gamma$-ray connection \citep{PhysRevLett.125.011101, 2026arXiv260220145M}, they generally require coupling with auxiliary components (such as RIAFs or external CR reservoirs) to cover the PeV range, which increases the model parameter space. Conversely, our model physically connects the low- and high-energy spectral components through the temporal evolution of a single source population. In the context of Seyfert galaxies, our model is highly complementary to the disk-corona scenario. Highly obscured, Compton-thick sources such as NGC 1068 and NGC 4151 likely have their neutrino fluxes dominated by the innermost optically thick corona, whereas such sources during their quiescent phases are also well described by the diffuse host-galaxy reservoir model presented here, and thus contribute to the DNB. Although we have adopted simplifications, such as a uniform active lifetime, the characteristic spectral features predicted by our model arise from the fundamental physics of energy-dependent CR transport under episodic AGN activity and are robust against minor parameter variations. Future Monte Carlo simulations incorporating a continuous distribution of AGN lifetimes and duty cycles will further refine these predictions.

% One may argue that the final result depends on the AGN X-ray luminosity function, which may not apply to all AGN types, and that simplified parameter choices such as a uniform active phase lifetime are adopted. However, the predicted spectral features rely only on CR evolution under AGN variability and are not model- or parameter-dependent. A more precise calculation using Monte Carlo simulations\textemdash generating individual AGN neutrino emission and summing their contributions\textemdash could be performed in future work, and we expect the main conclusion to remain unchanged.

Crucially, our model predicts distinct observational signatures that can be tested by future multi-messenger campaigns. At low redshifts, the quiescent-phase host galaxies are predicted to emit sub-TeV $\gamma$-rays, which is tentatively supported by recent \textit{Fermi}-LAT detections of GeV--TeV emissions from radio-quiet AGNs that exceed corona model predictions \citep{2025NatAs...9.1086L}. Stacking analyses of TeV-bright, radio-quiet AGNs will provide a critical test of this prediction. Furthermore, the spatial morphology of neutrino emission serves as a smoking gun to distinguish our model from core-dominated scenarios (e.g., the disk-corona model): corona models predict point-like neutrino emission localized at the SMBH, whereas our model predicts spatially extended emission tracing the gas distribution of the host galaxy. Resolving these spatial differences requires neutrino telescopes with angular resolutions approaching $0.01^\circ$. Under our quiescent-phase reservoir scenario, next-generation deep-sea or deep-ice neutrino observatories like NEON \citep{ZHANG2025103123} are projected to detect neutrino events from individual sources such as CGCG 420-015 at a rate of $\sim 0.5~{\rm yr^{-1}}$, enabling a $5\sigma$ discovery within approximately two months of operation and opening a new window into AGN physics and CR transport.

% The neutrino emission in our model may be accompanied by sub-TeV $\gamma$-rays at low redshift. The detection of $\gamma$-ray flux from \textit{Fermi} radio-quiet AGNs that exceeds the compact corona predictions in the energy range between several GeV and hundreds of GeV \citep{2025NatAs...9.1086L} shows tentative evidence for this prediction. Stacking analysis of neutrino emission from low-redshift TeV-bright radio-quiet AGNs can further test this scenario. Another key distinction between our model and previous ones lies in the neutrino emission sites, which lead to different spatial morphologies: neutrinos produced in scenarios such as the disk-corona model are expected to originate near the SMBH, whereas those from AGNs in the quiescent phase would be diffusely distributed throughout the host galaxy. This morphological difference calls for neutrino telescopes with extremely high angular resolution\textemdash on the order of $0.01^{\circ}$\textemdash to both localize sources and identify their physical origins. If the quiescent phase origin holds, a next-generation observatory such as NEON \citep{ZHANG2025103123, 2025heas.confE...3X} could detect neutrino events from CGCG 420-015 with a maximum rate of $\sim0.5$ yr$^{-1}$, requiring only about two months to reach $5\sigma$ significance.

\appendix
\section{Appendix}

\subsection{Magnetic and radiation field models in the jet and radio lobe}

The magnetic and radiation field environments within an active galactic nucleus (AGN) are essential for estimating particle acceleration and cooling timescales. In this work, we adopt a widely used magnetic field model based on flux conservation, in which the magnetic field strength scales with the distance $z$ from the supermassive black hole (SMBH) as \citep{1999agnc.book.....K, 2017SSRv..207....5R}
\begin{equation}
\label{eq-a1}
    B(z)=B_{0}\left(\frac{z_{0}}{z}\right)^{m}.
\end{equation}
Here, $B_{0}$ is the magnetic field strength at a distance $z_{0}$ from the SMBH. The parameter $z_{0}$ represents the jet launch region and is taken to be the characteristic scale of the AGN corona \citep{2017SSRv..207....5R}. The index $1\leq m\leq2$ is determined by the magnetic field topology \citep{1999agnc.book.....K} (for a purely toroidal magnetic field, $m=1$, while for a purely poloidal one, $m=2$). In this work, we set $m=1$; however, since the magnetic field in the core region is constrained to be on the order of Gauss \citep{PhysRevD.90.023007}, this choice has no significant impact on our results.

The radiation field model of an AGN is more complex than the magnetic counterpart. The total radiation field comprises contributions from the accretion disk, the corona, the broad-line region, the dust torus, the host galaxy bulge, the relativistic jet, and the cosmic microwave background. We adopt the canonical blazar radiation field model presented in \citep{10.1111/j.1365-2966.2009.15007.x} and adopt the energy density $U_{\rm{jet}}$ of jet emission as 
\begin{equation}\label{eq-a2}
    U_{\rm{jet}}\approx\frac{L_{\rm{jet}}}{4\pi\Gamma^{4}H_{j}^{2}c},
\end{equation}
where $L_{\rm{jet}}$ is the bolometric radiation luminosity of the jet \citep{PhysRevD.84.103007, J.Patrick_Harding_2012, PhysRevD.90.023007}, $\Gamma$ represents the bulk Lorentz factor of the jet, and $H_{j}$ denotes the comoving size of the emission region.

\subsection{Timescales for high-energy particles}

In our model, particles \textemdash including electrons and protons\textemdash are assumed to be accelerated primarily in the AGN core. They subsequently propagate through the jet and eventually form the radio lobe. Throughout this process, particles undergo acceleration via diffusive shocks, plasma turbulence, or both, as well as energy losses (cooling) due to synchrotron radiation, inverse Compton (IC) scattering, proton-proton ($pp$) interactions, photomeson production ($p\gamma$ interactions), and Bethe-Heitler pair production. We characterize these competing processes by their respective acceleration and cooling timescales, from which the energy spectra of nonthermal particles are derived.

We first consider the acceleration process. The timescale of first-order Fermi acceleration via diffusive shocks is given by \citep{PhysRevD.90.023007, 2017SSRv..207....5R} 
\begin{equation}\label{eq-a3}
    t_{\rm{acc-1}}(E)=\frac{E}{\eta ec^{2}B},
\end{equation}
where $\eta\leq1$ describes the acceleration efficiency. The timescale for second-order Fermi acceleration through plasma turbulence can be expressed as \citep{Kimura_2015, PhysRevD.100.083014, PhysRevLett.125.011101, 2021NatCo..12.5615K}
\begin{equation}\label{eq-a4}
    t_{\rm{acc-2}}(E)=\frac{1}{\zeta}\left(\frac{c}{v_{A}}\right)^{2}\frac{H_{a}}{c}\left(\frac{E}{ecBH_{a}}\right)^{2-q},
\end{equation}
when assuming gyro-resonant scattering through turbulence with a power spectrum of $P_{k}\propto k^{-q}$. The coefficient $\zeta$ is the ratio of the turbulent magnetic field strength to that of the non-turbulent component. The quantity $H_{a}$ represents the comoving size of the acceleration region.

For electrons, synchrotron radiation and IC scattering are the dominant energy-loss channels. Their cooling timescales are given by \citep{Zhang_2021}
\begin{align}\label{eq-a5}
    t_{\rm{syn-e}}(E)=\;&\frac{3m_{e}^{2}c^{3}}{4\sigma_{T}EU_{B}},\\
    \label{eq-a6}t_{\rm{IC-e}}(E)=\;&\frac{3m_{e}^{2}c^{3}}{4\sigma_{T}EU_{\rm{rad}}}\left(1+\frac{4E\epsilon_{0}}{m_{e}^{2}c^{4}}\right)^{3/2}.
\end{align}
Here, $m_{e}$ is the electron mass and $\sigma_{T}$ is the Thomson cross section. The term $U_{B}=B^{2}/(2\mu_{0})$, with $\mu_{0}$ the vacuum permeability, denotes the energy density of the magnetic field, while $U_{\rm{rad}}$ represents the energy density of radiation fields. The factor $[1+4E\epsilon_{0}/(m_{e}^{2}c^{4})]^{3/2}$ describes the Klein-Nishina effect, where $\epsilon_{0}=2.82k_{b}T_{\rm{rad}}$ is the typical photon energy of the target blackbody/graybody radiation field, with $k_{b}$ and $T_{\rm{rad}}$ being the Boltzmann constant and radiation temperature, respectively. 

Protons also undergo cooling through the above two channels, with their cooling timescales given by expressions similar to Eqs.~(\ref{eq-a5}) and (\ref{eq-a6}), scaled by a factor of $m_{p}^{4}/m_{e}^{4}$, where $m_{p}$ is the proton mass. Since $m_{p}$ is more than three orders of magnitude larger than $m_{e}$, the lifetime of nonthermal protons is expected to be significantly longer than that of nonthermal electrons in most scenarios. Three additional channels contribute to proton energy losses: $pp$ interactions, $p\gamma$ interactions, and Bethe-Heitler pair production. Their cooling rates take the following forms \citep{PhysRevLett.125.011101}
\begin{equation}\label{eq-a7}
    t_{\rm{pp-p}}^{-1}(E)=n_{p}\sigma_{\rm{pp}}(E)k_{\rm{pp}}c,
\end{equation}
and
\begin{align}\label{eq-a8}
    t_{\rm{p\gamma-p}}^{-1}(E)=\;&\frac{c}{2\gamma_{p}^{2}(E)}\int_{\bar{E}_{\rm{th}}}^{\infty}\mathrm{d}\bar{E}_{\gamma}\sigma_{\rm{p\gamma}}(\bar{E}_{\gamma})k_{\rm{p\gamma}}(\bar{E}_{\gamma})\bar{E}_{\gamma}\nonumber\\\;&\times\int_{\bar{E}_{\gamma}/[2\gamma_{p}(E)]}^{\infty}\mathrm{d}E_{\gamma}E_{\gamma}^{-2}n_{\gamma}(E_{\gamma}),\\
    \label{eq-a9}t_{\rm{BH-p}}^{-1}(E)=\;&\frac{c}{2\gamma_{p}^{2}(E)}\int_{2m_{e}c^{2}}^{\infty}\mathrm{d}\bar{E}_{\gamma}\sigma_{\rm{BH}}(\bar{E}_{\gamma})k_{\rm{BH}}(\bar{E}_{\gamma})\bar{E}_{\gamma}\nonumber\\\;&\times\int_{\bar{E}_{\gamma}/[2\gamma_{p}(E)]}^{\infty}\mathrm{d}E_{\gamma}E_{\gamma}^{-2}n_{\gamma}(E_{\gamma}).
\end{align}
In Equation~(\ref{eq-a7}), $n_{p}$ is the cold proton density, while $\sigma_{\rm{pp}}(E)$ and $k_{\rm{pp}}=0.5$ denote the cross section and inelasticity for $pp$ interactions. Here, we adopt the $\sigma_{\rm{pp}}(E)$ from \citep{PhysRevD.90.123014}. In Eq.~(\ref{eq-a8}), $\gamma_{p}(E)=E/(m_{p}c^{2})$ is the proton Lorentz factor, $\bar{E}_{\rm{th}}\approx145$ MeV is the threshold energy for photomeson production, and $\bar{E}_{\gamma}$ and $E_{\gamma}$ represent the photon energy in the proton rest frame and lab frame, respectively. The terms $\sigma_{\rm{p\gamma}}(\bar{E}_{\gamma})$ and $k_{\rm{p\gamma}}(\bar{E}_{\gamma})$ are the cross section and inelasticity for $p\gamma$ interactions, and $n_{\gamma}(E_{\gamma})$ describes the photon density per unit energy. As indicated by Eq.~(\ref{eq-a9}), the cooling rate for the Bethe-Heitler pair production is analogous to that of the $p\gamma$ interactions, with the only modifications being the lower limit of the outer integral changed from $\bar{E}_{\rm{th}}$ to $2m_{e}c^{2}$, and the cross section and inelasticity replaced by $\sigma_{\rm{BH}}(\bar{E}_{\gamma})$ and $k_{\rm{BH}}(\bar{E}_{\gamma})$, where we adopt the fitting formula given in \citep{1992ApJ...400..181C}. 

\begin{figure*}[!t]
    \centering
    \includegraphics[width=1.0\textwidth]{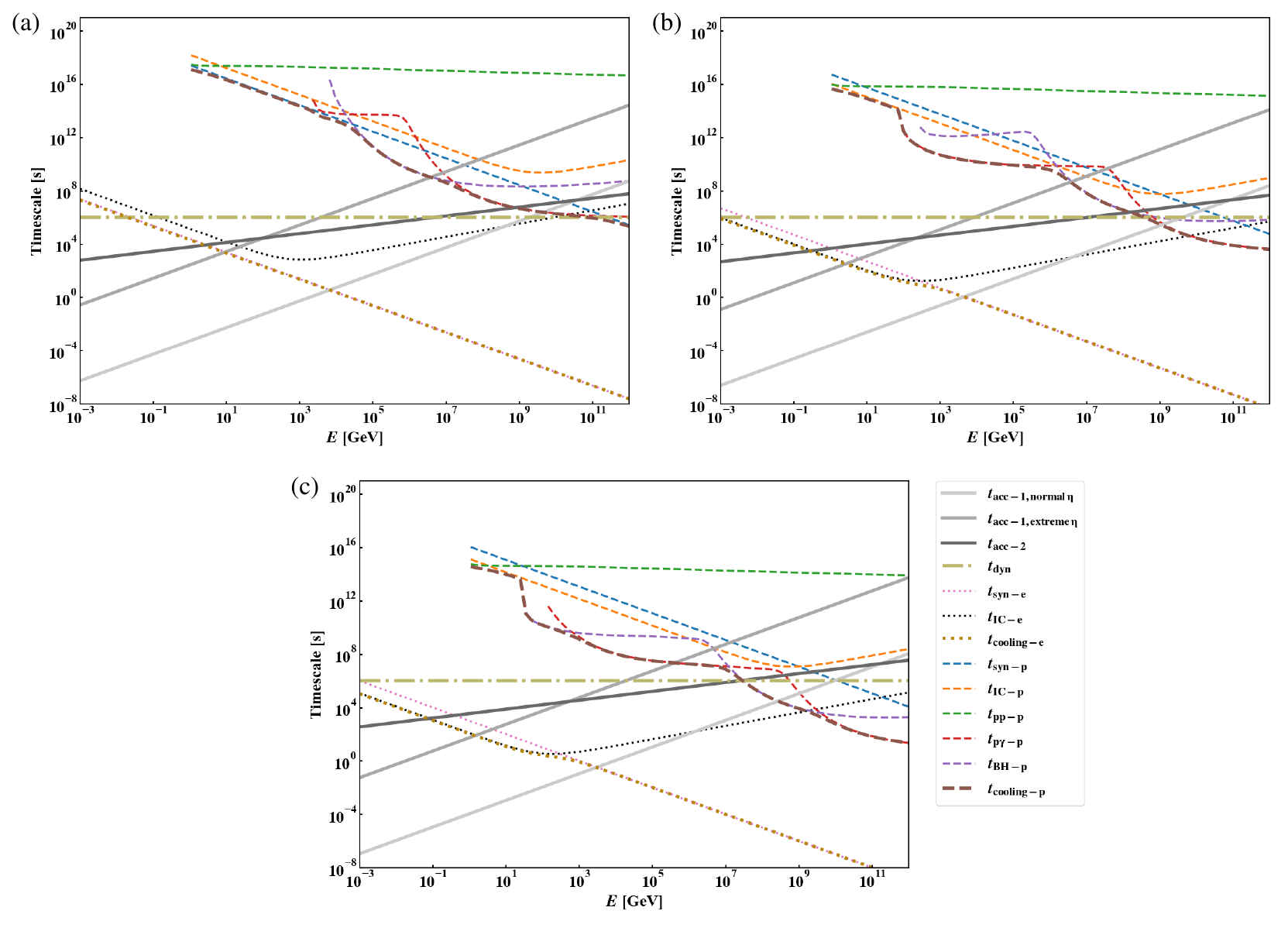}
    \caption{\label{fig:S1}Key timescales of high-energy particles in the AGN core region versus X-ray luminosity. Panels (a)-(c) correspond to $L_{X}=1.0\times10^{42}$, $1.0\times10^{44}$, and $1.0\times10^{46}$ erg s$^{-1}$, respectively. Electron timescales are computed over the energy range of $10^{-3}$ to $10^{12}$ GeV, while proton timescales cover $1$ to $10^{12}$ GeV. The absence of low-energy components for the $p\gamma$ and Bethe-Heitler cooling curves arises because the integrals in Eqs.~(\ref{eq-a8}) and (\ref{eq-a9}) vanish at those energies. The first-order Fermi acceleration timescales with a normal $\eta=0.05$ and an extreme $\eta=1.0\times10^{-7}$ are plotted as silver and gray lines, respectively.}
\end{figure*}

Using the magnetic and radiation field model presented above, we calculate the acceleration and cooling timescales for nonthermal particles in AGNs. Since particles are primarily accelerated in the core region\textemdash where the magnetic and radiation field environments around the SMBH are the most intense and complex\textemdash the associated timescales are expected to play a crucial role in shaping their spectral evolution. In Fig.~\ref{fig:S1}, we present the acceleration and cooling timescales for high-energy particles in the core region, along with the dynamical timescale $t_{\rm{dyn}}=1.0\times10^{6}$ s \citep{PhysRevD.90.023007}. The calculation is performed in the jet comoving frame; we adopt a scale height of $H_{c}=1.0\times10^{-5}$ kpc to estimate the magnetic field strength and the radiation field energy density in the core region, and assume $H_{j}=H_{a}=H_{c}$. For second-order Fermi acceleration, the efficiency parameter $\zeta$ is set to a plausible value of 0.05 \citep{PhysRevLett.125.011101}. For first-order Fermi acceleration, we consider two values of $\eta$: a typical value $\eta=0.05$\textemdash equal to $\zeta$ for direct comparison\textemdash and an extremely low value $\eta=1.0\times10^{-7}$, chosen to mimic the regime where both mechanisms yield comparable acceleration timescales for electrons.

As shown in Fig.~\ref{fig:S1}, at the same efficiency, first-order Fermi acceleration significantly outperforms second-order Fermi acceleration in both low- and high-energy bands, although its effectiveness diminishes more rapidly with increasing particle energy. When the two mechanisms produce comparable electron energy cutoffs, first-order Fermi acceleration is much less efficient than second-order Fermi acceleration at energies above TeV. Consequently, as discussed in the main text, under the assumption that electrons and protons share a common acceleration mechanism and given the constraints from the blazar sequence, our model favors second-order Fermi acceleration as the dominant process in the jet. For simplicity, we neglect first-order Fermi acceleration in our calculations, which does not imply that shocks are absent in the jet.

The timescales for high-energy particles in the collimation zone and radio lobe can be derived similarly, using different scale heights to estimate the magnetic field strength and radiation field energy density. To simplify the calculation, we adopt uniform scale heights of $H_{z}=0.1$ kpc for the collimation zone and $H_{l}=2.0$ kpc for the lobe. Note that $H_{c}$, $H_{z}$, and $H_{l}$ are parameters used solely in these estimates and do not represent the physical sizes of the corresponding regions. In addition, for the collimation zone, we set $t_{\rm{dyn}}$ to $1.0\times10^{10}$ s; for the lobe, we incorporate an adiabatic cooling timescale $t_{\rm{adia}}=5.26\times10^{14}$ s, equivalent to $5/3$ of the active phase timescale \citep{10.1111/j.1365-2966.2010.17982.x, refId6}.

\subsection{Diffuse neutrino and gamma-ray background modeling}

As shown in Fig.~\ref{fig:2}, our model successfully reproduces the convex diffuse neutrino background (DNB) observed by IceCube \citep{2gh9-d4q7} and predicts an accompanying diffuse $\gamma$-ray background that falls below the \textit{Fermi} isotropic diffuse $\gamma$-ray background \citep{Ackermann_2015}. This provides a novel perspective: the observed neutrino sky is the cosmological superposition of neutrino emission from AGNs at different evolutionary stages. The diffuse neutrino flux is derived from Eq.~(\ref{eq-6}), whereas the $\gamma$-ray flux can be obtained by a similar equation
\begin{align}\label{eq-a10}
	\Phi_{\gamma}(E_{\gamma})=\;&\frac{c}{4\pi H_{0}}\int_{0}^{z_{\rm{max}}}\frac{\mathrm{d}z}{\sqrt{(1+z)^{3}\Omega_{m}+\Omega_{\Lambda}}}\nonumber\\\;&\times\int_{0}^{T_{\rm{max}}} \frac{\mathrm{d}T}{T_{\rm{max}}}\int_{L_{X, \rm min}}^{L_{X, \rm max}}\mathrm{d}L_{X}\frac{\mathrm{d}\rho_{X}}{\mathrm{d}L_{X}}(z)\nonumber\\\;&\times\frac{L_{E'_{\gamma}}(E_{\gamma},z,T,L_{X})}{E'_{\gamma}}e^{-\tau_{\gamma}(E_{\gamma},z)}.
\end{align}
Here, $L_{E'_{\gamma}}$ is the $\gamma$-ray luminosity per unit energy of an individual source, which depends on the intrinsic $\gamma$-ray energy $E'_{\gamma}=(1+z)E_{\gamma}$ at redshift $z$. The index $\tau_{\gamma}(E_{\gamma},z)$ denotes the optical depth arising from extragalactic background light (EBL), which increases steeply with energy in the TeV range. In the calculation, we retrieve $\tau_{\gamma}(E_{\gamma},z)$ using the editable code and adopt the EBL model of \citep{Finke_2022}.

The neutrino ($\gamma$-ray) luminosity per unit energy $L_{E'_{\nu}}$ ($L_{E'_{\gamma}}$) of an individual AGN is derived from Eqs.~(\ref{eq-1})-(\ref{eq-2}) and (\ref{eq-5}), in combination with the widely used $p\gamma$ \citep{Hummer_2010} and $pp$ \citep{PhysRevD.74.034018} interaction models as well as the \texttt{GAMERA} toolkit \citep{2015ICRC...34..917H}. The specific form of the term $Q_{0}(t)$ in Eq.~(\ref{eq-2}) is \citep{2017SSRv..207....5R}
\begin{equation}\label{eq-a11}
	Q_{0}(t)=\begin{cases}
    \frac{L_{e}}{V\Gamma m_{e}c^{2}}&it_{c}<t\leq it_{c}+t_{f} \\
    \frac{0.01L_{e}}{V\Gamma m_{e}c^{2}}&it_{c}+t_{f}<t\leq (i+1)t_{c} 
\end{cases},
\end{equation}
for electrons and 
\begin{equation}\label{eq-a12}
	Q_{0}(t)=\begin{cases}
    \frac{L_{p}}{V\Gamma m_{p}c^{2}}&it_{c}<t\leq it_{c}+t_{f} \\
    \frac{0.01L_{p}}{V\Gamma m_{p}c^{2}}&it_{c}+t_{f}<t\leq (i+1)t_{c} 
\end{cases},
\end{equation}
for protons. Here, $V$ represents the physical volume of the corresponding region, $i\geq0$ indicates the number of flare cycles, $t_{c}=1$ yr is the full cycle duration, and $t_{f}=0.01$ yr denotes the flare duration. Since the flare electron and proton powers, $L_{e}$ and $L_{p}$, depend on the X-ray luminosity $L_{X}$ (see Eqs.~(\ref{eq-3})-(\ref{eq-4})) and $L_{X}$ is a more universal proxy for all AGNs compared to other radiation luminosities, we model $L_{E'_{\nu}}$ and $L_{E'_{\gamma}}$ as functions of $L_{X}$, and assume that AGNs with the same $L_{X}$ share uniform parameters. 

The value of $T$ determines whether an AGN is in the active or quiescent phase and also affects its proton spectrum. For AGNs in the active phase, neutrinos/$\gamma$-rays are produced in the core region, collimation zone, and radio lobe. In the core and collimation zone, we adopt steady-state proton spectra for calculation; in the lobe, the proton spectrum is determined by $T$, as high-energy CRs accumulate over time. For AGNs in the quiescent phase, as CRs\textemdash particularly those with higher energies\textemdash gradually escape from the host galaxy, the resulting neutrino and $\gamma$-ray emission depends critically on $T$. Even with a low diffusion coefficient of $1.0\times10^{28}(E/4\,\rm{GeV})^{0.29}$ cm$^{2}$ s$^{-1}$, PeV CRs would leave the host galaxy in tens of millions of years. Consequently, the dominant contribution of quiescent AGNs is concentrated in this timescale. 

\begin{figure*}[!t]
    \centering
    \includegraphics[width=0.75\textwidth]{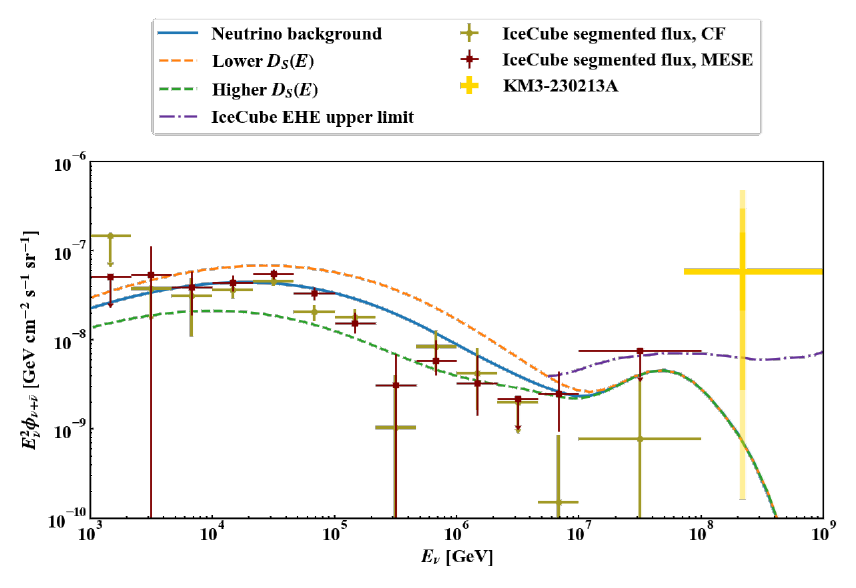}
    \caption{\label{fig:S2}Model-predicted diffuse neutrino flux per neutrino flavor for various diffusion coefficients. The blue solid line shows the result obtained with $D_{S}(E)=1.0\times10^{28}(E/4\,\rm{GeV})^{0.29}$ cm$^{2}$ s$^{-1}$, while the orange and green dashed lines correspond to $D_{S}(E)=5.0\times10^{27}(E/4\,\rm{GeV})^{0.29}$ and $2.0\times10^{28}(E/4\,\rm{GeV})^{0.29}$ cm$^{2}$ s$^{-1}$, respectively. The IceCube and KM3NeT data and upper limits \citep{PhysRevD.98.062003, 2gh9-d4q7, 2025Natur.638..376K} are also plotted for comparison, with the same conventions as in Fig.~\ref{fig:2}.}
\end{figure*}

Due to the relativistic jet motion, the neutrino and $\gamma$-ray emission from the core region and collimation zone is Doppler boosted \citep{2007Ap&SS.311..231K}. Here, we use a Doppler factor $\bar\delta$ averaged over the viewing angle $\theta$ and weighted by $\delta^{r}$ to evaluate the Doppler boosting for individual energies. The $\bar\delta$ is calculated following
\begin{align}\label{eq-a13}
    \bar\delta=\;&\frac{2}{\pi}\frac{\int_{0}^{\pi/2}\mathrm{d}\theta\delta(\theta)\delta(\theta)^{r}}{\int_{0}^{\pi/2}\mathrm{d}\theta\delta(\theta)^{r}}\nonumber \\=\;&\frac{2}{\pi}\frac{\int_{0}^{\pi/2}\mathrm{d}\theta[\Gamma(1-\beta\cos\theta)]^{-(r+1)}}{\int_{0}^{\pi/2}\mathrm{d}\theta[\Gamma(1-\beta\cos\theta)]^{-r}},
\end{align}
where $\Gamma$ is the bulk Lorentz factor of the relativistic jet and $\beta=(1-1/\Gamma^{2})^{1/2}$ represents the jet speed normalized to the speed of light. Moreover, the average flux enhancement is given by 
\begin{equation}\label{eq-a14}
    \bar{\delta^{r}}=\frac{2}{\pi}\int_{0}^{\pi/2}\mathrm{d}\theta[\Gamma(1-\beta\cos\theta)]^{-r}.
\end{equation}
The above calculations assume a uniform distribution of $\theta$, and we adopt $\Gamma=10$ and $r=2$ \citep{2007Ap&SS.311..231K}.

Using the methods described above, we obtain the results shown in Fig.~\ref{fig:2}; the key parameters adopted in the computation are given in Table~\ref{tab:1}. The most noteworthy point is the adoption of $D_{S}(E)$ lower than the Galactic value, which implies that AGN host galaxies serve as more efficient CR reservoirs than previously thought \citep{Tamborra_2014, Hooper_2016}. This emerges naturally from the IceCube observations when AGN variability is taken into account. In Fig.~\ref{fig:S2}, we show in detail the impact of $D_{S}(E)$ on the model-predicted DNB, by varying $D_{S}(E)$ by factors of 0.5 and 2 while keeping the scale height $H_{g}$ fixed. As shown in the figure, a lower $D_{S}(E)$ shifts the spectral break to higher energies and increases the flux intensity. Conversely, a higher $D_{S}(E)$ results in a spectral break at lower energies and a reduced intensity. Thus, the spectral break energy observed by IceCube is intrinsically linked to the flux intensity, both jointly determined by the diffusion coefficient\textemdash or more precisely, by the diffusive escape time of CRs. This further validates that the spectral features of the DNB originate from AGN variability, and that the observed break energy suggests that most AGN host galaxies confine CRs for longer timescales than the Milky Way.

The slower diffusion of CRs during the quiescent phase is supported by several observational and theoretical studies. It is evident that the radio lobes generated by AGN activity can persist for $\sim10^{6}$--$10^{7}$ years after the power supply from the AGN is switched off \citep{refId0, Sanderson_2024}, indicating that the turbulent magnetic field can endure long after the central engine becomes inactive and continue to impede the diffusive escape of CRs from the galaxy. In addition, studies have pointed to a "fossil" phase of AGNs \citep{10.1093/mnras/stt1150}, during which metal ions remain out of ionization equilibrium for up to $\sim10^{8}$ years after AGN activity ceases. This suggests that AGN activity can influence its environment on timescales significantly exceeding its active phase lifetime. CR diffusion is found to be suppressed in star-forming regions of galaxies \citep{Semenov_2021, Pandey_2024}. Given that AGN feedback can act as a trigger for star formation \citep{10.1111/j.1365-2966.2012.22074.x, Dugan_2014}, this process could also lead to a reduced value of $D_{S}(E)$. 

\subsection{Individual source modeling}

In recent years, IceCube has detected several promising point source candidates, including the transient source TXS 0506$+$056 \citep{doi:10.1126/science.aat1378} and the five brightest nearby Seyfert galaxies\textemdash NGC 1068, NGC 4151, NGC 7469, CGCG 420-015, and the Circinus Galaxy \citep{Carpio_2026}. We apply our model to these sources. As shown in Fig.~\ref{fig:4}, TXS 0506$+$056 can be interpreted in the active phase scenario, whereas NGC 7469, CGCG 420-015, and the Circinus Galaxy are well described by the quiescent phase scenario.  

For TXS 0506$+$056, after computing the electron and proton spectra in the core region, collimation zone, and radio lobe, we adopt \texttt{GAMERA} \citep{2015ICRC...34..917H} to compute the photon spectrum and the Sim-B code \citep{Hummer_2010} to derive the neutrino and $\gamma$-ray spectrum in each region. The calculation takes as input the photon spectral energy distributions (SEDs) and the proton spectra of the three regions, and yields the $E_{\nu}^{2}$-weighted neutrino and $\gamma$-ray densities per unit energy per unit time for each region. To obtain the flux observed at Earth, these outputs are multiplied by the volumes of the corresponding regions, divided by $4\pi d_{L}^{2}$, where $d_{L}$ is the luminosity distance, and then summed. For the core and collimation zone, the neutrino and $\gamma$-ray energy are boosted by the Doppler factor $\delta$ and the flux is enhanced by Doppler boosting $\delta^{r}$. For individual source modeling, the Doppler factor is set to $\delta\approx\Gamma=10$, while the spectral index is fixed at $r=2$, consistent with the DNB modeling. 

\begin{figure*}[!t]
    \centering
    \includegraphics[width=0.6\textwidth]{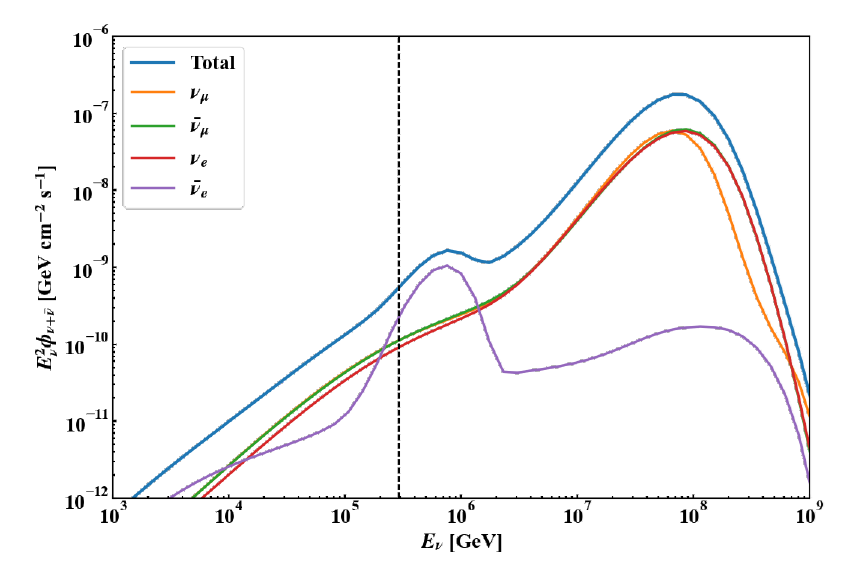}
    \caption{\label{fig:S3}Flavor composition of the model-predicted neutrino flux from TXS 0506$+$056. The blue solid line represents the all-flavor neutrino flux, whereas the orange, green, red, and purple solid lines correspond to the components of muon neutrino, muon antineutrino, electron neutrino, and electron antineutrino, respectively. The black dashed line indicates the most probable energy of IceCube-170922A associated with TXS 0506$+$056 \citep{doi:10.1126/science.aat1378}.}
\end{figure*}

As discussed, the neutrino event IceCube‑170922A associated with TXS 0506$+$056 is coincident with the neutron‑decay‑generated peak predicted by our model. Here, we present the detailed flavor composition of the modeled neutrino flux of TXS 0506$+$056 in Fig.~\ref{fig:S3}. The electron antineutrino peak at $\sim700$ TeV may account for the IceCube-170922A event at $\sim290$ TeV, given the non-negligible uncertainties in both the event energy \citep{doi:10.1126/science.aat1378} and the $L_{X}$ of TXS 0506$+$056 \citep{Fiorillo_2025}. This suggests a physical scenario in which the electron synchrotron/IC radiation field interacts with high-energy protons near the energy threshold via hyperon
resonances, producing a neutron peak. The subsequent decay of these neutrons yields an electron antineutrino peak whose energy aligns with that of IceCube-170922A. One of these electron antineutrinos then oscillates into a muon neutrino and is eventually detected by IceCube. 

Seyfert galaxies are also considered promising candidates for astrophysical neutrino sources \citep{Murase_2024, Carpio_2026}. As discussed, we modeled five representative nearby Seyfert galaxies\textemdash NGC 1068, NGC 4151, NGC 7469, CGCG 420-015, and the Circinus Galaxy. Since most Seyfert galaxies are radio-quiet\textemdash which is more consistent with the quiescent phase\textemdash and the primary associated neutrino fluxes are concentrated in the energy range from 10 TeV to 100 TeV \citep{Abbasi_2026a, Abbasi_2026b}, also characteristic of the quiescent phase, we adopt the quiescent phase scenario. To model their neutrino and $\gamma$-ray emission, we first calculate the electron and proton spectra in the core region, collimation zone, and radio lobe with the same procedure as in the active phase. We then take the CR spectrum in the lobe at the end of the active phase ($t=1.0\times10^{7}$ yr) as the initial condition for the quiescent phase. The time dependent CR spectrum in the quiescent phase is thereby obtained by numerically solving Eq.~(\ref{eq-5}). Using this spectrum along with the widely adopted $pp$ interaction model \citep{PhysRevD.74.034018} and \texttt{GAMERA}, we calculate the neutrino and $\gamma$-ray flux from the five sources.

For NGC 7469, CGCG 420-015, and the Circinus Galaxy, our model successfully reproduces their neutrino spectra and $\gamma$‑ray fluxes consistent with the \textit{Fermi}-LAT upper limits, adopting physically plausible parameters. The results are presented in Fig.~\ref{fig:4} (b)-(d). According to Table~\ref{tab:2}, the expected lifetimes of the three Seyfert galaxies are concentrated in 2--$3\times10^{7}$ yr, comparable to the diffusive escape timescale of PeV CRs. The gas density of 1 cm$^{-3}$ applied to NGC 7469 is identical to that used in the DNB modeling; however, for CGCG 420-015 and the Circinus Galaxy, this value is reduced to 0.3 cm$^{-3}$ and 0.05 cm$^{-3}$, respectively. The value adopted for CGCG 420-015 remains in line with the typical value for spiral galaxies \citep{vallee1997observations}, while the Circinus Galaxy, though several times larger than the Milky Way in size, has a comparable total gas mass \citep{10.1046/j.1365-8711.1999.02057.x}, which reasonably explains its lower gas density. These parameter choices are therefore physically justified.

\begin{figure*}[!t]
    \centering
    \includegraphics[width=1.0\textwidth]{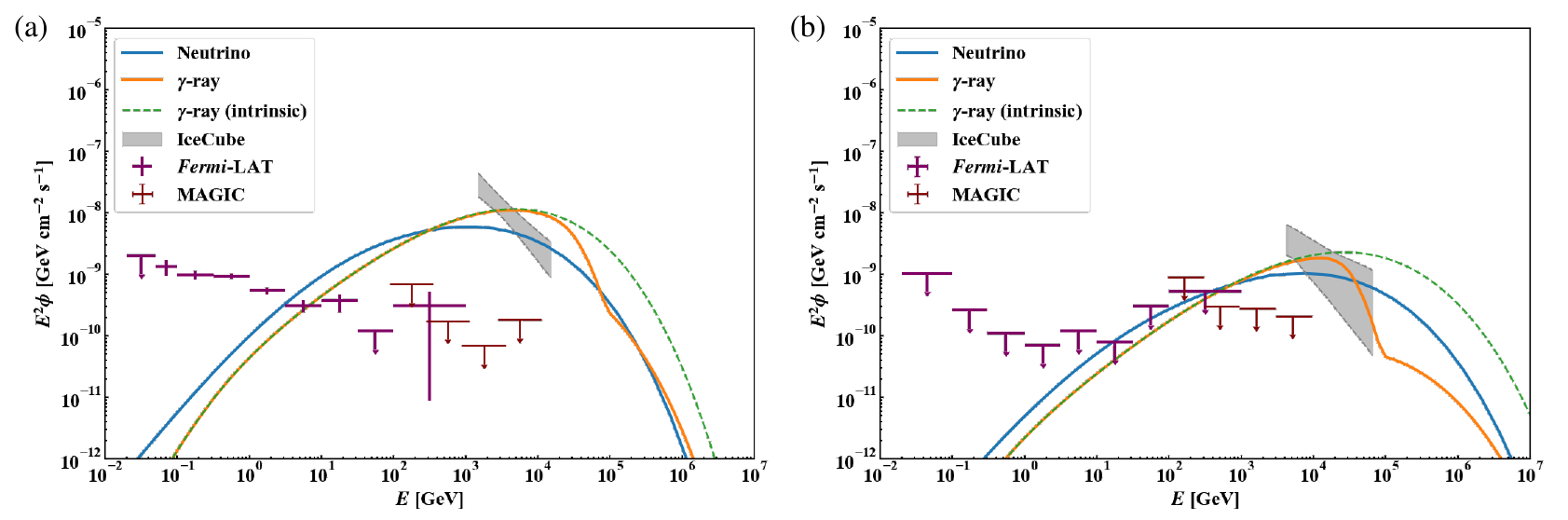}
    \caption{\label{fig:S4}Model-predicted neutrino flux per neutrino flavor and the accompanying $\gamma$-ray spectrum for NGC 1068 and NGC 4151, compared with both neutrino and $\gamma$-ray observations. Panel (a) shows the results for NGC 1068. The IceCube butterfly data is taken from \citep{doi:10.1126/science.abg3395}, alongside the \textit{Fermi}-LAT \citep{Ajello_2023} and MAGIC \citep{Acciari_2019} data. Panel (b) gives the results for NGC 4151, with the IceCube butterfly data from \citep{Abbasi_2025}, and the \textit{Fermi}-LAT and MAGIC data from \citep{Murase_2024} and \citep{refId7}, respectively.}
\end{figure*}

For NGC 1068 and NGC 4151, Fig.~\ref{fig:S4} presents the modeled neutrino and $\gamma$-ray spectra of the two sources alongside IceCube \citep{doi:10.1126/science.abg3395, Abbasi_2025}, \textit{Fermi}-LAT \citep{Ajello_2023, Murase_2024}, and MAGIC observations \citep{Acciari_2019, refId7}, with the primary adopted parameters summarized in Table~\ref{tab:S1}. Stringent $\gamma$-ray upper limits have been set by \textit{Fermi}-LAT and MAGIC in the 100 GeV to 10 TeV range, exhibiting little correlation with the neutrino spectra, implying that the $\gamma$-ray emission is likely to be absorbed.  The quiescent phase scenario assumes neutrino production throughout the host galaxy, which cannot provide a $\gamma$-ray-opaque environment unless extremely high gas densities are invoked. Additionally, since both NGC 1068 and NGC 4151 are low-redshift AGNs, EBL absorption can only significantly attenuate $\gamma$-rays above 100 TeV. Thus, our model may not be appropriate to simultaneously account for the neutrino and $\gamma$-ray emission of these two sources. For NGC 4151, models favoring neutrino production near the $\gamma$-ray-opaque core region\textemdash such as the disk-corona \citep{Carpio_2026, 2026arXiv260727658L} or outflow–cloud interaction \citep{2025arXiv251106707M}\textemdash offer plausible interpretations. However, for NGC 1068, the aforementioned models have difficulty explaining the low-energy portion around 1 TeV of its neutrino spectrum. Furthermore, the spectral shape of NGC 1068 deviates from the DNB trend observed before the lower-energy break (see Fig.~5 in \citep{Abbasi_2026a}). This underscores the need for GeV-band measurements of the DNB to clarify the contribution from NGC 1068. If an additional break appears at tens of GeV with a flux substantially higher than that of NGC 1068, it would imply a large population of NGC 1068–like sources. Conversely, if the DNB at lower energies is primarily contributed by NGC 1068, this would imply that NGC 1068 is a peculiar outlier that warrants further, more in-depth investigation.

\renewcommand\arraystretch{1.25}
\begin{table*}[!t]
  \centering
  \caption{\label{tab:S1}Primary parameters adopted for modeling NGC 1068 and NGC 4151}
  % \begin{ruledtabular}
  \begin{tabular}{ccc}
  \hline
     & NGC 1068 & NGC 4151 \\ \hline
    $L_{x}$ [erg s$^{-1}$] & $3\times10^{43}$ \citep{Testagrossa_2026} &  $5\times10^{42}$ \citep{Testagrossa_2026} \\
    $d_{L}$ [Mpc] & 10.1 \citep{Testagrossa_2026} & 15.8 \citep{Testagrossa_2026} \\
    $T$ [yr] & $1\times10^{8}$ & $5\times10^{7}$ \\
    $n_{p}$ [cm$^{-3}$] & 1 & 1 \\
    $q_{\rm{rel}}q_{\rm{jet}}$ & 0.5\% & 0.5\% \\
    $L_{p}/L_{e}$ & 0.015 & 0.015 \\
    \hline
  \end{tabular}
  % \end{ruledtabular}
\end{table*}

\subsection{Detectability of point sources with next-generation neutrino observatories}

In this section, we briefly discuss the prospect of detecting point sources using the NEutrino Observatory in the Nanhai (NEON). NEON is a next-generation neutrino observatory proposed for deployment in the South China Sea \citep{ZHANG2025103123}. Its full detection array, arranged in a Fibonacci spiral tiling, will have a physical volume of $\sim10$ km$^{3}$ and consist of over 1000 strings, each carrying 18 digital optical modules (DOMs). The spacing between strings is 100 m and the vertical separation between adjacent DOMs on the same string is 40 m. Benefiting from its large volume, NEON is expected to achieve an angular resolution of $0.1^{\circ}$ at 100 TeV \citep{ZHANG2025103123}, which is crucial for precisely localizing sources and reducing background contamination. 

\begin{figure*}[!t]
    \centering
    \includegraphics[width=0.6\textwidth]{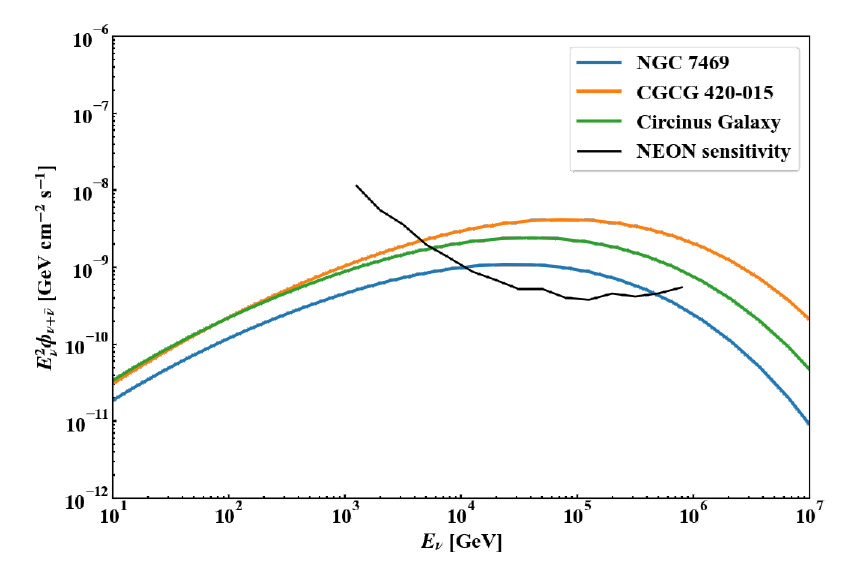}
    \caption{\label{fig:S5}Expected point source neutrino flux versus NEON sensitivity. The blue, orange, and green lines show the model-predicted neutrino flux of NGC 7469, CGCG 420-015, and the Circinus Galaxy, respectively, identical to those in Fig.~\ref{fig:4} (b)-(d). The black line represents the $3\sigma$ sensitivity of NEON for a deployment depth of 3500 m.}
\end{figure*}

As elaborated previously, our model can be used to interpret the neutrino emission from TXS 0506$+$056, NGC 7469, CGCG 420-015, and the Circinus Galaxy. Among these, the latter three are expected to be persistent sources and thus more favorable targets for detection. Assuming a quiescent phase origin, we compare the model-predicted neutrino spectra of these sources with the $3\sigma$ sensitivity curve of NEON in Fig.~\ref{fig:S5}. As shown in the figure, all three sources lie within the detection reach of NEON in the 10–100 TeV band. CGCG 420‑015 stands out as the most promising candidate, with its neutrino emission expected to be observable from several TeV up to above 1 PeV. At these high energies, the effective area of NEON exceeds 1 km$^{3}$ \citep{ZHANG2025103123} while the atmospheric background becomes negligible \citep{PhysRevD.92.023004}. These features make NEON promising to identify point sources similar to CGCG 420-015 in the future and reveal the origin of high-energy astrophysical neutrinos.

\begin{figure*}[!t]
    \centering
    \includegraphics[width=1.0\textwidth]{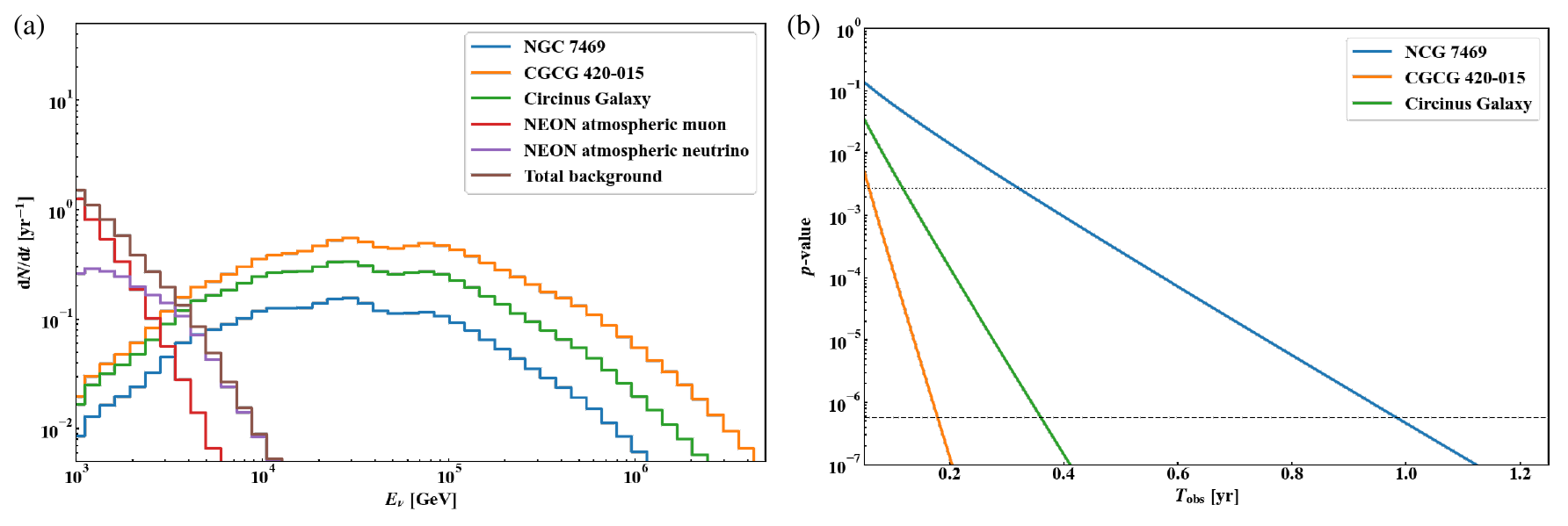}
    \caption{\label{fig:S6}Detection prospects for NGC 7469, CGCG 420-015, and the Circinus Galaxy with NEON. Panel (a) shows the muon neutrino event rates from the three sources along with the background components and their total in different energy bins. Panel (b) presents the corresponding $p$-values as a function of observation time. The black dotted and dashed lines indicate the $p$-values required to reach $3\sigma$ and $5\sigma$ significance, respectively.}
\end{figure*}

Taking NGC 7469, CGCG 420-015, and the Circinus Galaxy as highly promising targets for NEON, we estimate their muon neutrino event rates in different energy bins using \citep{2025heas.confE...3X}
\begin{equation}\label{eq-a15}
	\frac{\mathrm{d}N(E_{\nu})}{\mathrm{d}t}=\int_{E'_{\nu1}}^{E'_{\nu2}}\mathrm{d}E'_{\nu}\frac{A_{\nu}^{\rm eff}}{4\pi d_{L}^{2}}\frac{L_{E'_{\nu}}}{E'_{\nu}}(E_{\nu}),
\end{equation}
where $E'_{\nu1}$ and $E'_{\nu2}$ define the boundaries of the energy bin, and $A_{\nu}^{\rm eff}$ is the muon neutrino effective area of NEON \citep{ZHANG2025103123, 2025heas.confE...3X}. The corresponding background within the angular resolution of NEON (see Fig.~9 in \citep{ZHANG2025103123}) is also calculated. The background consists of atmospheric muons, atmospheric neutrinos, and the diffuse neutrino flux. According to our estimation, the diffuse neutrino flux is significantly lower than the other components; therefore, we neglect its contribution to the total background. The intrinsic atmospheric muon flux for NEON is simulated using CORSIKA, and the intrinsic atmospheric neutrino flux is based on the Honda model \citep{PhysRevD.92.023004}. For atmospheric muons, we expect only $10^{-4}$ to survive after discrimination, while 55\% of the neutrino signal is expected to be retained \citep{ZHANG2025103123}. For atmospheric neutrinos, a preliminary self-veto is applied. All the above procedures are performed assuming a deployment depth of 3500 m. Figure~\ref{fig:S6} (a) presents our results. Taking advantage of the high angular resolution and large effective area, NEON gives a high signal-to-noise ratio in the range from TeV to PeV and a maximum event rate of $\sim0.5$ yr$^{-1}$ from CGCG 420-015 at approximately 30 TeV. The $p$-value for detecting the three sources as a function of observation time, computed using the method in \citep{Kheirandish_2021}, is shown in Fig.~\ref{fig:S6} (b). For CGCG 420-015, a $5\sigma$ detection could be achieved in about two months, while even for the weakest target, NGC 7469, the same significance level would be reached within one year.

In summary, for NEON in its current configuration, it would be highly sensitive to promising nearby Seyfert galaxies under the quiescent phase assumption. A $5\sigma$ detection of the brightest source, CGCG 420‑015, is expected to be achievable within two months of observation using NEON, and even for the faintest target, NGC 7469, the same significance can be reached in one year. A stacking analysis with IceCube-Gen2 and KM3NeT could further reduce this timescale.

%
% ****** End of file apssamp.tex ******

\acknowledgments

This work is supported by the National Natural Science Foundation of China (NSFC) grants 12261141691. 

% \paragraph{Note added.} This is also a good position for notes added
% after the paper has been written.

% Bibliography

%% [A] Recommended: using JHEP.bst file
\bibliographystyle{JHEP}
\bibliography{biblio}

%% or
%% [B] Manual formatting (see below)
%% (i) We suggest to always provide author, title and journal data or doi:
%% in short all the informations that clearly identify a document.
%% (ii) please avoid comments such as "For a review'', "For some examples",
%% "and references therein" or move them in the text. In general, please leave only references in the bibliography and move all
%% accessory text in footnotes.
%% (iii) Also, please have only one work for each \bibitem.

% \begin{thebibliography}{99}

% \bibitem{a}
% Author,
% \emph{Title},
% \emph{J. Abbrev.} {\bf vol} (year) pg.

% \bibitem{b}
% Author,
% \emph{Title},
% arxiv:1234.5678.

% \bibitem{c}
% Author,
% \emph{Title},
% Publisher (year).

% \end{thebibliography}
\end{document}